\documentclass[12pt,preprint]{aastex}
\slugcomment{submitted to ApJ}
\def\kxt{\xi}
\def\kyt{\eta}
\def\omt{\Omega}
\def\zt{\zeta}
\def\pt{\varphi}
\def\vxt{u_x}
\def\vyt{u_y}
\def\vt{{u_z}}

\def\d1{{\rho_{1}}}
\def\rf{{\rm F}}
\def\half{\tiny{1/2}}
\def\quota{\tiny{1/4}}
\def\thalf{\tiny{3/2}}
\def\fhalf{\tiny{5/2}}
\def\shalf{\tiny{7/2}}

 \shortauthors{LEE AND HONG}
 \shorttitle{PARKER INSTABILITY IN A SELF-GRAVITATING MAGNETIZED GAS DISK}
\begin{document}
\title{Parker Instability in a Self-Gravitating Magnetized Gas Disk:\\
I. Linear Stability Analysis}
\author{Sang Min Lee\altaffilmark{1,4}
and S. S. Hong\altaffilmark{2,3,5}}
 \altaffiltext{1}{Supercomputing Center, KISTI, Daejeon, 305-600,
Korea}
 \altaffiltext{2}{Astronomy Program, SEES, Seoul National
 University, Seoul 151-742, Korea}
 \altaffiltext{3}{Department of Infrared Astrophysics, ISAS/JAXA, Sagamihara, Kanagawa, Japan}
 \altaffiltext{4}{e-mail: smlee@kisti.re.kr}
 \altaffiltext{5}{e-mail: sshong@astroism.snu.ac.kr}

\begin{abstract}
To be a formation mechanism of such large-scale structures as
giant molecular clouds $\,$(GMCs) and HI superclouds, the
classical Parker instability driven by external gravity has to
overcome three major obstacles: The convective motion accompanying
the instability generates thin sheets than large condensations.
The degree of density enhancement achieved by the instability is
too low to make dense interstellar clouds. The time and the length
scales of the instability are significantly longer and larger than
the estimated formation time and the observed mean separation of
the GMCs, respectively. This paper examines whether a replacement
of the driving agent from the external to the self gravity might
remove these obstacles by activating the gravitational instability
in the Galactic ISM disk. Self-gravitating, magnetized, gas disk
bound by a hot halo medium is subject to a Parker-type instability
driven by the self gravity, the usual Jeans instability and the
convection. If the disk is thicker than a certain critical value,
the Parker-type instability due to perturbations of undular mode
assists the Jeans gravitational instability to form large-scale
structures. Under the external gravity, the convection triggered
by interchange mode perturbations is to grow most rapidly; under
the self gravity, however, the Jeans gravitational instability can
have a growth rate higher than the limiting growth rate of the
convective instability. The self gravity can thus suppress the
convective motions, and a cooperative action of the Jeans and the
Parker instabilities can remove all the obstacles confronting the
classical version of the Parker instability. The mass and mean
separation of the structures resulting from the odd-parity undular
mode solution are shown to agree better with the HI superclouds
than with the GMCs. We briefly discuss how inclusions of the
external gravity and cosmic rays would modify behaviors of the
odd-parity undular mode solution.
\end{abstract}

\section{INTRODUCTION}
In seminal studies of Parker in mid 1960s, he proved that
magnetized gas disk under the influence of a constant
gravitational acceleration becomes unstable to long wavelength
perturbations (Parker 1966, 1967). The instability named after
Parker was once thought to be the formation mechanism of giant
molecular cloud complexes$\,$(GMCs) in the Galaxy$\,$(Parker 1966;
Mouschovias 1974; Appenzeller 1974; Mouschovias et al. 1974; Blitz
\& Shu 1980; Shibata \& Matsumoto 1991; Handa et al. 1992). This
notion was based mostly on the results of stability analysis done
in two dimensions, where only undular mode can be taken into
account. Many subsequent analyses done in three
dimensions$\,$(eg., Zweibel \& Kulsrud 1975; Hughes \& Proctor
1988), however, called our attention to the interchange mode,
which may quickly drive the system convectively unstable. Since
the convection grows faster for shorter wavelength perturbations,
it tends to shred the Galactic ISM disk into thin sheets of low
density before the undular mode of long wavelength perturbations
collects enough material to form well-defined large-scale
structures near the midplane$\,$(Ass\'eo et al. 1978;
Lachi\`eze-Rey et al. 1980). To accept the classical Parker
instability as the formation mechanism of such galactic structures
as the GMCs or the HI superclouds, one needs means to control the
disruptive tendency of the interchange mode. The observed density
in GMCs is about a few tens of hydrogen molecules per cubic
centimeter. On the other hand, three-dimensional MHD simulations
of the Parker instability in uniform and non-uniform external
gravities render to the system only a factor 2 enhancement in
density over its initial equilibrium value, which is far too low
to generate interstellar clouds of even moderate density$\,$(Kim
et al. 1998; Franco et al. 2002). This low degree of density
enhancement is one of the most serious obstacles the Parker
instability should overcome to be the GMC formation mechanism.

Under canonical conditions of the Galactic ISM disk$\,$(cf.,
Spitzer 1978), the classical Parker instability driven by a
uniform external gravity has the minimum growth time of
$\sim$10$^{8}$ years and the corresponding length scale
$\sim$1.6$\,$kpc, which are clearly incompatible with the GMC
observations. It was in 1990's that non-uniform models of gravity
were tried out as a more realistic driving agent than the uniform
one$\,$(Giz \& Shu 1993; Kim \& Hong 1998). By introducing a
non-uniform external gravity, whose acceleration increases
linearly with the vertical distance, $z$, from the midplane up to
$z\simeq$ 500$\,$pc and stays constant there on, one may reduce
the time and length scales nominally to the degree compatible with
the GMCs$\,$(Hong \& Kim 1997; Kim \& Hong 1998). However, the
reduction of scales was not enough to give candidacy of the GMC
formation mechanism back to the classical Parker instability,
because the reduced scale values are relevant for activities in
the regions of high Galactic altitudes, where vertical
acceleration of the chosen external gravity becomes strongest but
the GMCs do not generally reside.

The undular mode of perturbations triggers the Parker instability
preferably in those disks, where magnetic field strength decreases
rapidly with vertical distance, because it is easy for such disk
to develop strong magnetic buoyancy $\,$(Parker 1955; Hughes \&
Proctor 1988).  Under the uniform external gravity the field and
density decrease with $z$ exponentially; while under the
non-uniform external gravity they follow roughly a Gaussian
function of $z$. Therefore, it is easier to trigger the Parker
instability under the non-uniform gravity than the uniform one.
This is how a replacement of the external driving agent from the
uniform to the non-uniform gravity has brought to the system a
significant reduction in the time scale. However, the reduction of
time scale doesn't necessarily mean that dense condensations form
near the midplane within a reasonably short period of time. It
only means a quick development of diffuse chaotic features high
above the central plane.

This line of reasoning led us to replace the driving agent of the
Parker instability from the external gravity originated mostly
from the Galactic stars to the internal one of the ISM itself. We
expect the self gravity will ease the problem of low degree of
density enhancement the classical Parker instability faced. The
self gravity would prevent the Galactic ISM disk from developing
convective activities, while it would also activate the Jeans
gravitational instability preferably in the disk midplane. In this
paper we will examine time, length and mass scales involved with
the Parker and the Jeans instabilities in a magnetized gas disk
under the self gravity.

There have been many studies on the effect of magnetic fields upon
the Jeans gravitational instability. Chandrasekhar \&
Fermi$\,$(1953) examined development of the gravitational
instability in a uniformly magnetized, plane-parallel gas layer.
They showed that the existence of uniform fields doesn't change
the critical wavelength for onset of the Jeans instability but
only reduces its growth rate. When non-uniform field is
considered, the critical wavelength does change and the gas layer
becomes more unstable than the uniformly magnetized
case$\,$(Stod\'olkeiwicz 1963; Nakano 1988). Nakamura et
al.$\,$(1991) showed that self-gravitating, magnetized gas disk is
subject to the Parker-type and the Jeans instabilities
simultaneously. Since all of these investigations were done in the
plane defined by the vertical and the magnetic field directions,
they couldn't address the question whether the self gravity could
put a curb upon flaring motions of the convection.
Elmegreen$\,$(1982) is the first who seriously examined possible
roles of the self gravity in modifying the classical picture of
the Parker instability. His stability analysis was, in principle,
of the three-dimensional nature. However, he integrated the basic
MHD equations over the whole extent of the disk, which made the
analysis two-dimensional in reality.

In a more recent study, Nagai et al.$\,$(1998) made a fully
three-dimensional stability analysis onto a uniformly magnetized,
self-gravitating gas disk. However, the Parker instability could
not be triggered in their systems, because the magnetic fields
were not vertically stratified. Instead they examined how the disk
thickness might influence the alignment of filamentary structures
with respect to the magnetic field direction. Chou  et
al.$\,$(2000) carried out an extensive analysis on non-uniformly
magnetized, self-gravitating gas layers, for some cases including
the external gravity as well. Their linear analysis was further
extended to nonlinear, three-dimensional MHD simulations. In their
systems, however, the interchange mode didn't have a chance to
drive convective motions, because only isothermal perturbations
were applied to the background equilibrium which was in the same
isothermal condition. And contrary to our objectives their main
concern was on parsec-scale structures in the interstellar medium.
We took adiabatic perturbations in one of our earlier
three-dimensional stability analyses and saw a possibility of
controlling the convection by the self gravity$\,$(Lee \& Hong
1999; Lee et al. 2004).

In principle the Parker instability in the Galactic ISM disk
should be studied under simultaneous influence of the external and
the self gravities. In this study we take the self gravity as the
sole driving agent of the instability and examine what
modifications it could bring to the classical picture of the
Parker instability. This paper is organized as follows: In \S$\,$2
we will construct an initial equilibrium configuration for the
self-gravitating, magnetized, isothermal gas disk, and then
perform linear stability analysis onto it. In \S$\,$3 we will
interpret the resulting dispersion relations in terms of the
convective, the Parker and the Jeans instabilities. In \S$\,$4 we
will assess whether introduction of the self gravity as the
driving agent might ease the major problems confronting the
classical Parker instability. In the same section the resulting
scales of the maximum growth rate will be compared with the
observations of HI superclouds and GMCs. In the last section we
will conclude the paper with some discussions. Appendix A details
on coefficients of the linearized perturbation equations, and in
Appendix B stable wave modes are isolated. Marginal stability
analysis is carried out in Appendix C.

\section{BASIC FORMULATIONS}

\subsection{Initial Equilibrium Model}
We model the local part of the Galactic ISM disk as a
self-gravitating, magnetized, isothermal gas layer, which is
initially in magnetohydrostatic equilibrium and bound by a hot
halo at $z$ = $\pm$$z_{\rm a}$. Unperturbed magnetic field $\vec
B_{{\mathrm{o}}}$ = $\left[\,0, B_{\mathrm{o}}(z), 0\right]$ has
azimuthal component only and its strength varies with vertical
distance.

In such disk the vertical structure is described by the equation
of magnetohydrostatic equilibrium,
\begin{equation}
{{d}\over{dz}}\left[P_{\mathrm{o}}(z)
+{{B_{\mathrm{o}}(z)^2}\over{8 \pi}}\right] =-\rho_{\mathrm{o}}(z)
{{d \psi_{\mathrm{o}}(z)}\over{dz}},
\end{equation}
where $P$, $\psi$, and $\rho$ denote gas pressure,
self-gravitational potential, and gas density, respectively. The
potential is to be obtained from the Poisson equation,
\begin{equation}
{{d^2}\over{{dz}^2}} \psi_{\mathrm{o}}(z) = 4 \pi G
\rho_{\mathrm{o}}(z)\,,
\end{equation}
and the gas pressure is related to density as
\begin{equation}
P_{\mathrm{o}}(z)=c_{\rm s}^2 \rho_{\mathrm{o}}(z)
\end{equation}
with the isothermal sound speed, $c_{\rm s}$, taken to be equal to
the $rms$ velocity of interstellar `coudlets.' In this study
cosmic rays are ignored; instead, we will focus on the role of
self gravity. The ratio, $\alpha ={{B^2_{\mathrm{o}}(z)}/{8\pi
P_{\mathrm{o}}(z)}}$, of magnetic to gas pressure is kept constant
of $z$ for the initial equilibrium. Equations (1) through (3) then
give us the equilibrium stratifications of density and magnetic
field
\begin{equation}
\rho_{\mathrm{o}}(z) = \rho_{\mathrm{o}}(0)
\,{\mathrm{sech}^2}(z/H)\, ,
\end{equation}
\begin{equation}
{B_{{\mathrm{o}}}(z)} = B_{{\mathrm{o}}}(0) \,{\mathrm{sech}}(z/H)
\, ,
\end{equation}
and the resulting gravitational potential
\begin{equation}
\psi_{{\mathrm{o}}}(z) = 4 \pi G \rho_{{\mathrm{o}}}(0) H^2
\,{\mathrm{ln[\,cosh}}(z/H)],
\end{equation}
where scale height $H$ is defined by
\begin{equation}
H = c_{\rm s} \sqrt{{1+\alpha}\over{2\pi G
\rho_{{\mathrm{o}}}(0)}}.
\end{equation}

\subsection{Linearization}
The linearized MHD equations for a self-gravitating magnetized gas
disk are given by
\begin{equation}
{\partial \rho_1\over\partial t}  +  \nabla\cdot (\rho_{\mathrm
o}\vec v) = 0\,,
\end{equation}
\begin{equation}
\rho_{\mathrm o}{\partial \vec v \over \partial t}= -\nabla P_{1}
+ {1\over 4\pi}\,(\nabla\times{\vec B}_{1})\times{\vec B}_{\mathrm
o} + {1\over 4\pi}\,(\nabla\times{\vec B}_{\mathrm o})\times {\vec
B}_{1} - \rho_1\nabla\psi_{\mathrm o} - \rho_{\mathrm
o}\nabla\psi_{1}\,,
\end{equation}
\begin{equation}
{\partial\vec B_{1}\over\partial t}= \nabla\times (\vec v
\times{\vec{B}_{\mathrm o}})\,,
\end{equation}
\begin{equation}
{\partial P_1 \over \partial t}  +\gamma {P_{\mathrm{o}}}
\nabla\cdot{\vec v} +(\vec v\cdot\nabla){P_{\mathrm{o}}} = 0\,,
\end{equation}
and
\begin{equation}
{\nabla}^2 \psi_1 = 4 \pi G \rho_1\,,
\end{equation}
where all the symbols have their usual meanings and the ones with
subscript $1$ denote their perturbed quantities.  We take
adiabatic relation $P = \kappa \rho^{\gamma}$ as the equation of
state for the perturbed material [Eq. (11)].

Seeking solutions of the form, $ q_1(x,y,z,t) = q_1(z)\,\exp{[\,i(
k_x x +  k_y y - \omega t)]}$, we recast the linearized equations
in dimensionless forms as
\begin{equation}
\omt\,{\rho_{1} \over \rho_{\mathrm o}} =\, {\kxt}\,\vxt \,+\,
{\kyt}\,\vyt \,+\, 2\,i\,\Theta\,\vt \,-\, i\,{d\vt\over d\zt}\,,
\end{equation}
\begin{equation}
\omt\,\vxt =\,{\kxt\over {1 + \alpha}}\,{{P_1}\over P_{\mathrm o}}
            \,-\,{{2 \alpha\,\kyt }\over 1 + \alpha}\,{B_{1x}\over B_{\mathrm o}}
            \,+\,{{2 \alpha\,\kxt }\over 1 + \alpha}\,{B_{1y}\over B_{\mathrm o}}
            \,+\,\kxt\,\pt\,,
\end{equation}
\begin{equation}
\omt\,\vyt =\,{\kyt \over 1 + \alpha}\,{P_{1}\over P_{\mathrm
o}}\,\, - \,i\,{2\alpha\,\Theta\over 1 + \alpha}\,{B_{1z}\over
B_{\mathrm o}}
            \,+\, \kyt\,\pt\, ,
\end{equation}
\begin{displaymath}
i\,\omt\,\vt =\,{1\over 1 + \alpha}\,{d\over d\zt}\,{P_{1}\over
P_{\mathrm o}}
         \,-\,{{2\Theta}\over 1 + \alpha}\,{P_{1}\over{P_{\mathrm o}}}
         -\,i\,{2\,\alpha\,\kyt\over 1+\alpha}\,{{B_{1z}}\over B_{\mathrm o}}
            \,+ \,{2\alpha\over 1 + \alpha}\,{d\over d\zt}
              {B_{1y}\over B_{\mathrm o}}
\end{displaymath}
\begin{displaymath}
\hskip 9truemm -\,{4\alpha\Theta\over 1 + \alpha} \, {B_{1y}\over
B_{\mathrm o}}\, +\, {{d \pt}\over{d\zt}}\, +\,
2\Theta\,{\rho_{1}\over \rho_{\mathrm o}}\,,
\end{displaymath}
\begin{equation}
\omt{{B_{1x}}\over{B_{\mathrm o}}} = - \,\kyt\,\vxt\,,
\end{equation}
\begin{equation}
\omt{{B_{1y}}\over{B_{\mathrm o}}} = -\,i\,{d\vt\over d\zt}+
i\,\Theta\, \vt + {\kxt}\,\vxt\,,
\end{equation}
\begin{equation}
\omt{{B_{1z}}\over{B_{\mathrm o}}} = \,-\,\kyt\,\vt\,,
\end{equation}
\begin{equation}
\omt\,{P_{1}\over P_{\mathrm o}}\, = \,2i\,\Theta \vt
                       +   \gamma\,(\kxt\,\vxt + \kyt\,\vyt)
                       - i \gamma {d\vt\over d\zt}\,,
\end{equation}
and finally
\begin{equation}
{d^{\,2}\pt\over d\zt^{2}}\,-\,(\kxt^2+\kyt^2)\,\pt = \,2\,(1 -
\Theta^2)\,{\rho_{1}\over \rho_{\mathrm o}}\,.
\end{equation}
Here we have taken $H$, $\sqrt{1+\alpha}$ $c_{\rm s}$ $(\equiv V)$
and $V/H$ as length, velocity and frequency units, respectively:
$\kxt$ = $k_x \,H$, $\kyt$ = $k_y\,H$, $\zt$ = $z/H$, $\vxt$ =
$v_x/V$, $\vyt$ = $v_y/V$, $\vt$ = $v_z/V$, $\pt$ = $\psi_1/V^2$,
$\omt$ = $\omega\,(V/H)^{-1}$, and $\Theta\equiv\tanh\zt$.
Eliminating all the variables in favor of $u_z$ and $\pt$, we
combine equations (13) through (20) into the following two
second-order ordinary differential equations:
\begin{equation}
A_{2}{d^{\,2}\over d\zt^{2}}\,\vt
 \,+\, A_{1}{d\over d\zt}\,\vt \,+\, A_{0}\,\vt \,=\,
B_{1} {d\over d\zt}\,\pt + {B_0}\, \pt
\end{equation}
and
\begin{equation}
C_{2}{d^{\,2} \over d\zt^{2}}\,{\pt}
 \,+\, C_{0}\, \pt \,=\,
D_{1} {d\over d\zt}\,\vt + D_{0}\,\vt\,.
\end{equation}
Coefficients $A_0$, $A_1$, $A_2$, $B_0$, $B_1$, $C_0$, $C_2$,
$D_0$, and $D_1$ are given in Appendix A.

\subsection{Boundary Conditions and Numerical Method}
To fix boundary conditions we will follow Goldreich \& Linden-Bell
(1965) and Tomisaka \& Ikeuchi (1983). Let us suppose that the
upper boundary of the disk is displaced from $z$ = $\pm$$z_a$ by a
small amount $\delta z = s\,\exp{[\,i\,(k_{x} x + k_{y} y - \omega
t)]}$ with $s \ll H$, which will then give us the vertical
component of perturbation velocity at the boundary as
\begin{equation}
v_z(z_a) = -i\, \omega\, \delta z.
\end{equation}
An application of the Gauss flux theorem to the perturbed boundary
gives us
\begin{equation}
-k \psi_1 (z_a) - {{d \psi_1}\over{d z}}\biggr|_{z_a-} =\, 4 \pi G
\rho_{\mathrm o} (z_a) \delta z,
\end{equation}
since $\psi_1(z_a+)$ satisfies the Laplace equation in the
rarefied halo.

A third condition comes from pressure continuity. It is the sum of
the gas and the magnetic pressures that should be kept continuous
at the boundary. By integrating $z$-component of the MHD momentum
equation from $z_a$ to $z_a+\delta z$, one may have the total
pressure at the perturbed boundary as
\begin{eqnarray}
P(z_{a} + \delta z) \,+\, {1\over 8\pi} B^{2}|_{z_{a} + \delta z}
\,=\, P(z_{a}) \,+\, {1\over 8\pi} B^{2}|_{z_{a}} \,-\,
\rho_{\mathrm o}(z_{a})\,{d \psi_{\mathrm o}\over d
z}\biggr|_{z_{a}}\,\delta z  \,+\, O({\delta z}^{2}). \nonumber
\end{eqnarray}
Therefore, the total boundary pressure,
$P_{\mathrm{tot}}^{\mathrm{disk}}$, calculated from the disk side
becomes
\begin{equation}
P_{\mathrm{tot}}^{\mathrm{disk}}(z_a+\delta z) = P_{\mathrm
o}(z_a)+P_{1}(z_a) +{1\over{8\pi}} B_{\mathrm o}^2\biggr|_{z_a}
+{1\over{4\pi}} B_{\mathrm o} B_{1y}\biggr|_{z_a}-\rho_{{\mathrm
o}}(z_a) {{d \psi_{\mathrm o}}\over{d z}}\biggr|_{z_a} \delta z
\end{equation}
to first order of $\delta z$. For the same boundary pressure,
$P_{\mathrm{tot}}^{\mathrm{halo}}$, but calculated from the halo
side, we may have
\begin{equation}
P_{\mathrm{tot}}^{\mathrm{halo}}(z_a+\delta z) = P_{\mathrm
o}(z_a)+ {1\over{8\pi}} {B_{\mathrm
o}}^2\biggr|_{z_a}+{1\over{4\pi}}B_{\mathrm o}
B_{1y}^{\mathrm{halo}}\biggr|_{z_a},
\end{equation}
where perturbation of the halo gas pressure
$P_{1}^{\mathrm{halo}}(z_a)$ has been ignored, as the halo medium
is thought to be very hot. Also ignored is the gravitational
potential term, because the halo is filled with a rarefied medium.
By equating $P_{\mathrm{tot}}^{\mathrm{disk}}$ to
$P_{\mathrm{tot}}^{\mathrm{halo}}$ at the boundary, we finally
have the pressure continuity condition as
\begin{equation}
P_1(z_a) +{1\over{4\pi}} B_{\mathrm o} B_{1y}\biggr|_{z_a}-
\rho_{\mathrm o}(z_a) {{d \psi_{\mathrm o}}\over{d
z}}\biggr|_{z_a} \delta z  = {1\over{4\pi}} B_{\mathrm o}
B^{\mathrm{halo}}_{1y}\biggr|_{z_a}.
\end{equation}

Now $B_{1y}^{\mathrm{halo}}$ has to be specified. From the
induction equation [Eq. (10)] we obtain
\begin{equation}
{\vec B}_{1} = -{\,k_{y}\over \omega}\,v_{x} B_{\mathrm o}\,{\hat
e}_{x} - \left[{i\over \omega}\,{{\partial}\over{\partial
z}}\,(v_{z} B_{\mathrm o}) - {\,k_{x}\over \omega}\,v_{x} B_{0}
\right]\,{\hat e}_{y}
               -{\,k_{y}\over \omega}\,v_{z} B_{\mathrm o}\,{\hat e}_{z}.
\end{equation}
On the other hand, no current density is expected in the halo,
because matter density there is extremely low. Therefore, ${\vec
B}_1^{\mathrm{halo}}$ can be given by magnetic potential $\phi_m$
simply as
\begin{equation}
\vec{B}^{\mathrm{halo}}_{1} = B_{{\mathrm o}} \nabla\phi_{m}
\end{equation}
with
\begin{equation}
\phi_{m} = {\mathrm A}\,\exp{(i k_x x + i k_y y - k z
)}\enskip{\mathrm{and}}\enskip k^2=k_x^2+k_y^2.
\end{equation}
Equation (28) tells us that the vertical component of the magnetic
field becomes $ B_{1z}^{\mathrm{disk}} = -\left(k_y/\omega\right)
v_{z} B_{\mathrm o}$ at the disk side boundary. From the halo
side, Equation (29) shows $B_{1z}^{\mathrm{halo}}=-k B_{\mathrm
o}\phi_m$. By equating $B_{1z}^{\mathrm{disk}}$ to
$B_{1z}^{\mathrm{halo}}$, we have $\phi_{m} = \left(k_y/{k
\omega}\right) v_z$ and the azimuthal field component
$B_{1y}^{\mathrm{halo}}\,=\, i ({k_y^2}/ k\omega)\,v_{z}
B_{\mathrm o}.$ And its substitution for $B_{1y}^{\mathrm{halo}}$
in equation (27) finally gives the pressure continuity condition
as:
\begin{equation}
P_1(z_a) \,+\, {1\over{4\pi}}\,B_{\mathrm o} B_{1y}\biggr|_{z_a}
\,-\, \rho_0(z_a)\, {{d \psi_{\mathrm o}}\over{d z}}\biggr|_{z_a}
\delta z \,=\, i\, {{B_{\mathrm o}^2}\over{4\pi}}\biggr|_{z_a}\,
{{k_y^2}\over{k\omega}}\,
 v_z(z_a).
\end{equation}

In terms of the dimensionless variables the upper boundary
conditions [Eqs. (23), (24), and (31)] take the following forms:
\begin{equation}
-i\Omega\, \delta\zeta = \vt(\zt_a),
\end{equation}
\begin{equation}
-k\pt(\zt_a)-{{d\pt}\over{d\zt}}\biggr|_{\zt_a} = 2
\,{\mathrm{sech}}^2(\zt_a)\, \delta\zeta ,
\end{equation}
and
\begin{displaymath}
{1\over2}\gamma\, i\Omega\, \mathcal{D}\, {\mathrm{cosh}}^2\zt_a
{{d ^2 \pt}\over{d \zt^2}} \biggr|_{\zt_a}
-\,i\,\Omega\left({1\over2}\gamma\, \mathcal{D} k^2
{\mathrm{cosh}}^2 \zt_a
       -\,2 \alpha \kxt^2\Omega^2 \right)\pt(\zt_a)
\end{displaymath}
\begin{displaymath}
-\,2\,(1+\alpha)\Theta i\Omega\, \mathcal{D} \delta \zeta
+\,2\alpha \left(\Omega^2-{{2\alpha}\over{1+\alpha}}\kyt^2\right)
\left(\Omega^2-{{\gamma}\over{1+\alpha}}\kyt^2\right) {{d\vt}
\over {d\zt}}\biggr|_{\zt_a}
\end{displaymath}
\begin{equation}
+\left[2\Theta(\gamma-1)
\,\mathcal{D}\,+\,{{2\alpha\kyt^2}\over{k}}\mathcal{D}
-2\Theta\alpha\left\{\Omega^4-\left({{\gamma-2}\over{1+\alpha}}\kxt^2
              +{{2\alpha+\gamma}\over{1+\alpha}}\kyt^2\right)\Omega^2
      +{{2\alpha\gamma}\over{(1+\alpha)^2}}\kyt^2
      k^2\right\}\right]\vt(\zt_a)= 0,
\end{equation}
where
\begin{eqnarray}
z_a/H = \zt_a  \enskip{\mathrm{and}}\enskip  \mathcal{D} =
\Omega^4-{{2\alpha+\gamma}\over{1+\alpha}}k^2 \Omega^2
   +{{2\alpha\gamma}\over{(1+\alpha)^2}}\kyt^2k^{2}.\nonumber
\end{eqnarray}

At the midplane, parity conditions should be specified: If
symmetric (odd--parity) solution is sought for vertical velocity
$u_z$, it should satisfy
\begin{equation}
{{d\pt}\over{d\zt}}\biggr|_{\zt=0}=0
\enskip{\mathrm{and}}\enskip\vt(\zt=0)=0
\end{equation}
at the midplane; while anti-symmetric (even--parity) one requires
\begin{equation}
\pt(\zt=0)=0
\enskip{\mathrm{and}}\enskip{{d\vt}\over{d\zt}}\biggr|_{\zt=0}=0.
\end{equation}
The odd-parity solutions are mirror symmetric with respect to the
midplane, and the even-parity ones are of the midplane crossing
type.

Along with the upper-boundary continuity and the midplane parity
conditions, equations (21) and (22) comprise an eigenvalue problem
for $\Omega^2$. For given perturbation wave vector ($\xi$, $\eta$)
we take a trial value for $\Omega^2$, and integrate the equations
from $\zeta$ = $\zeta_a$ to $\zeta$ = 0 with the fourth-order
Runge-Kutta method. If the resulting solution satisfies the chosen
parity condition at $z$ = 0, the trial is adopted as the
eigenvalue. If not, we take another trial and perform the same
integration until the solution fulfills the midplane parity
condition.

The resulting growth rate squared, $-\Omega^2$, is illustrated as
a function of $\nu_x=\xi\tanh \zeta_a$ and $\nu_y=\eta\tanh
\zeta_a$ (not of $\xi$ and $\eta$), which will keep the slope of
dispersion relation the same in limits $\nu_x \rightarrow$ 0 or
$\nu_y \rightarrow$ 0, regardless of the disk thickness (Elmegreen
\& Elmegreen 1978). If we notice total column density is
2$\rho_{\mathrm o}(0)\times H$ for unbound isothermal disk and
2$\rho_{\mathrm o}(0)\times H \tanh \zeta_a$ for halo-bound one,
we may call $\nu_x$ and $\nu_y$ normalized wave numbers that are
{\it effective} for the halo-bound disk. The three parameters
($\alpha$, $\gamma$, $\zeta_a$) characterize the disk system and
the normalized effective wave numbers ($\nu_x, \nu_y$) do the
perturbation.

\section{CONVECTION, PARKER AND JEANS INSTABILITIES}

Whenever a need arises to differentiate the undular-mode
instability driven by external gravity from that by self gravity,
we will use the term {\it classical} Parker, for which both odd-
and even-parities are applicable. In the context of self gravity,
we will call the instability triggered by the same undular mode
but even-parity perturbations as simply the Parker or the {\it
pure} Parker instability, while reserving the term Parker-Jeans
for the odd-parity undular mode. On the other hand, the
instability triggered by short wavelength perturbations of the
interchange mode will be called broadly as convection. Long
wavelength perturbations of the same mode may of course activate
the gravitational instability. In this section we will examine how
the convection, the pure Parker instability, and the Jeans
gravitational instability compete or collaborate with each other
in forming or de-structuring large-scale structures.

\subsection{Undular Mode}
To have a comparison standard we first calculate the dispersion
relation for the pure gravitational instability by taking $\alpha$
= 0, $\nu_x$ = 0 and imposing the odd-parity condition. For
several cases of disk thickness, Figure 1a illustrates how the
normalized growth rate of the gravitational instability changes
with the azimuthal effective wave number. In general, the Jeans
instability grows faster in thicker disk. However, once it is
thicker than about 4 times the scale height, the growth rate
becomes almost independent of the disk thickness.

With the same set of system parameters except for $\alpha$ = 1.0,
we calculate $-\Omega^2$ again as a function of $\nu_y$. The
resulting dispersion relations for odd- and even-parity solutions
are shown, in Figure 1b, by solid and dotted lines, respectively.
Because even-parity cannot activate the gravitational instability
$\,$(Simon 1965; Elmegreen \& Elmegreen 1978), by simply changing
the parity condition from even to odd, we may easily differentiate
role of the Parker instability from that of the Jeans in the
dispersion relation. The dotted lines in Figure 1b are, therefore,
for the Parker instability only; while the solid ones are for the
case where both the Parker and the Jeans instabilities operate
simultaneously. This led us to coin the term {\it Parker-Jeans}
especially for the odd-parity undular mode.

A comparison of the solid lines in Figures 1a and 1b indicates
that the presence of magnetic fields in moderately thick disks
boosts the growth rate of the gravitational instability
significantly. The Parker and the Jeans instabilities interact
constructively in the disks thicker than, say, one half the scale
height; while in the disks thinner than that, the same magnetic
fields now add rigidity or extra pressure to the gas, thereby
hindering growth of the gravitational instability. With marginal
stability analysis (see Appendix C) we have determined the
critical thickness, over which the two instabilities interact
constructively or destructively: It is 0.549$\,H$ for a moderately
magnetized ($\alpha\simeq$ 1), isothermal ($\gamma$ = 1) disk. As
can be seen from the dotted lines in Figure 1b, growth rate of the
Parker instability depends on thickness more sensitively than that
of the Jeans instability. The dotted line almost coinciding with
the horizontal axis is for $\zeta_a$ = 1, and the dispersion
relation of even-parity becomes hardly noticeable for
$\zeta_a\,\le\,$ 1.0. Comparison of the dotted (even-parity) and
solid (odd-parity) lines in Figure 1b suggests that a marked
increase can be achieved in the growth rate of the Parker
instability by triggering the gravitational instability. This
opens up a possibility of alleviating the time scale problem of
the classical Parker instability.

To understand how the magnetic field strength modifies the
dispersion relation for the undular mode, we fix the disk
thickness at 5$\,H$ and vary $\alpha$ over 0, 1, 2, 3, and 4. The
resulting $-\Omega^2$ is plotted, in Figure 2, against the
effective azimuthal wave number. As before the solid and dotted
lines are for odd- and even-parity solutions, respectively. It is
clear from the figure that the maximum growth rate of the Parker
instability increases with field strength (Notice difference in
the ordinates between Figures 1 and 2.). Under an influence of
vertical acceleration, the undulated field lines give the magnetic
buoyancy a chance to unload the gas material off their shoulders
towards the midplane valleys. As the field strength increases, so
does the buoyancy, and hence the system becomes highly unstable to
the perturbations of undular mode. The resulting escalation of the
Parker instability then boosts development of the gravitational
instability. The maximum growth rate of the odd-parity solutions,
therefore, increases significantly with $\alpha$. The critical
wave number for marginal stability is known to increase with
$\alpha$ as $(1 + 2\alpha)^{1/2}$ (Stod\'olkiewicz 1963), which
also indicates the constructive nature of Parker and Jeans
interaction.

We now turn our attention to the remaining system parameter,
$\gamma$, which was introduced as a measure of material rigidity.
To trigger the classical Parker instability under external
gravity, {\it effective} ratio of the specific heats for the
perturbed material should be less than a certain critical value,
$\gamma_{\rm c}$. Many authors have shown $\gamma_{\rm c} = 1 +
\alpha$ under external gravity$\,$(Newcomb 1961; Parker 1966;
Gilman 1970; Kim \& Hong 1998). In a recent study Lee$\,$(2002)
proved that the critical value takes the same 1 + $\alpha$ under
the influence of self gravity as well. To somewhat dramatize the
effect of $\gamma$ on the growth rate we decided to fix $\alpha$
at a rather small value 0.1 and examined the dispersion relation
with $\gamma$ varying 0.68 through 1.1. The solid and dotted lines
in Figure 3 are for the odd- and even-parity solutions,
respectively. For soft materials the dispersion curve exhibits two
maxima: The gravitational instability is responsible for the peak
occurring at smaller wave number and the Parker is for the other
at larger one. As $\gamma$ increases, height of the Parker peak
decreases. This is because, as the material becomes harder, the
magnetized gas system becomes more stable to the undular mode
perturbation. If $\gamma$ exceeds its critical value, the Parker
instability cannot be triggered at all. Owing to a heightening
activity of the Parker instability in softer media, the Jeans peak
becomes taller as $\gamma$ decreases.

For most cases shown in Figure 3, the dotted lines of even-parity
match smoothly with the solid lines of odd-parity. For somewhat
large $\gamma$'s, however, the even-parity perturbations of
undular mode grow a little faster than the odd-parity ones. Under
self gravity the growth rate difference between the even- and
odd-parity solutions is too small to render any significant
consequences to the spatial distribution of the resulting
condensations with respect to the midplane. In the case of
classical Parker instability, however, the even-parity
perturbations grow substantially faster than the odd-parity ones.
Therefore, the condensations, if they were generated by the
classical Parker instability, would place their maximum density
points on the northern and southern sides of the midplane
alternatively$\,$(Kim et al. 2000).

Figure 3 suggests that the undular mode of odd-parity triggers
both the Jeans and the Parker instabilities in the $\gamma$ = 0.68
disk, for example. The two most rapidly growing perturbations in
the disk have $\lambda_y\simeq$ 8.6$\,H$ for the Jeans instability
and $\lambda_y\simeq$ 3.5$\,H$ for the Parker case. On the other
hand, the perturbations of the same undular mode with the same
$\lambda_y\simeq$ 8.6$\,H$ but of even-parity would turn off the
gravitational instability and drive the system to develop only the
Parker. We have shown, in Figures 4a, 4b and 4c, eigen-solutions
for the cases with $\lambda_y\simeq$ 8.6$\,H$ of odd-parity, with
$\lambda_y\simeq$ 3.5$\,H$ of odd-parity, and with
$\lambda_y\simeq$ 8.6$\,H$ of even-parity, respectively. In the
left panel, the color changes from dark red to white as density
increases, and the numbers marked on each contours represent
logarithm of the density perturbation, $\rho_{1}(y, z)$, measured
in units of the initial equilibrium density, $\rho_{\rm o}(z=0)$,
at the midplane. In the right hand panels, contours in thin solid
lines are for the magnetic fields and the ones in thick lines
delineate the halo boundary. Small arrows inside the panels
represent velocity vectors, a big arrow on top of frame {\it a}
and {\it b} corresponds to two times the sound speed, and that of
frame {\it c} does just one.

An action of the self gravity is apparent from the velocity
pattern in Figure 4a. In the region of inflated field lines the
magnetic buoyancy drives matter to rise against the self gravity;
while in the magnetic valley the self gravity does the driving in
an opposite direction and collects material towards the center. We
see motions taking place everywhere in the $yz$-plane. Even in the
midplane one may witness a converging motion, which is made
possible by the self gravity and cannot be expected under the
external gravity. As a final product of the Jeans instability the
system will bear out a dense clump centered at Y $\simeq$ 4.3.
Velocity vectors in Figure 4b exhibit a pattern markedly different
from the ones in Figure 4a. Magnitude of the velocity vectors
quickly diminishes towards central plane, and they hardly exhibit
any motions in the midplane. The magnetic field lines in low
altitudes maintain their initial configuration almost intact.
Features of the perturbed density, the magnetic fields and the
velocity patterns in Figure 4b are all very similar to the
corresponding cases of the classical Parker instability that is
driven by an external gravity of linear model$\,$(Lee 1993). As a
final product of the Parker instability the system will bear out
clumps centered at Y $\simeq$ 1.7, 5.2 and 8.5. The iso-density
contours for odd-parity solutions in Figures 4a and 4b demonstrate
dense clumps formed right in the midplane. The deformation of
boundary is closely related to the enhanced density near the
midplane$\,$(Elmegreen \& Elmegreen 1978). A rise and a fall of
the boundary surface from its initial equilibrium height $z_{a}$ =
5$\,H$ accompany a fall and a rise in the midplane density,
respectively.

Eigen-solutions of the even-parity shown in Figure 4c make a
marked contrast to the odd-parity solutions in Figures 4a and 4b.
The upper and lower boundaries make the rise-and-fall motions in
unison. Active motions are apparent only in high above the
midplane. These features are very much reminiscent of the Parker
instability driven by a linear model of external gravity$\,$(Lee
1993). It is quite clear that the classical Parker instability may
not be the formation mechanism of such large-scale structures as
GMCs or HI superclouds. The GMCs are attributes of the Galactic
midplane, while the Parker instability generates actions at high
altitudes. To have such dense structures as GMCs in the disk
central plane, we need to invoke the self gravity not only to
drive the Parker undular mode but also to trigger the Jeans
gravitational instability.

As far as forming large-scale structures is concerned, the {\it
odd-parity} undular mode deserves our special attention, because
it will drive the system to develop the Jeans gravitational
instability and the Parker instability simultaneously. If the two
instabilities work constructively, they would generate dense
condensations in the central plane. On the other hand, the undular
mode perturbations of {\it even-parity} drive the system to
develop only the Parker instability, whose activities are confined
mostly to high altitude regions, where GMCs ar not generally
found. The question then is whether the resulting condensations of
the odd-parity solutions are safe from disruption by convective
motions.

\subsection{Interchange Mode}
To construct dispersion relation for the {\it pure} interchange
mode, we fix azimuthal wave number at zero ($\nu_y$ = 0) and
calculate the growth rate as a function of effective radial wave
number $\nu_x$. The results for the odd-parity solutions are
shown, in Figure 5, for the disks of a moderately magnetized
($\alpha$ = 1), self-gravitating, isothermal gas medium. As was
seen from Figure 1a for the non-magnetized gas disk, the maximum
growth rate generally increases with disk thickness. However, for
the disks thicker than $\sim$5$\,H$, the dispersion relations for
different thickness are hardly distinguishable from each other.
Once the upper boundary is placed beyond a certain distance from
the midplane, development of the gravitational instability in the
densest part of the system doesn't depend much on where the
boundary is located. In disks thinner than $\sim$$\,H$/3, the
growth rate again becomes independent of the thickness. In such
thin halo-bound disks, there is practically no gradient in density
over vertical distance, and the gas acts like an incompressible
fluid (Goldreich \& Lynden-Bell 1965; Elmegreen \& Elmegreen 1978;
Lubow \& Pringle 1993; Nagai et al. 1998; Lee 2002). A sizeable
difference exists in the maximum growth rate between
non-magnetized (see Figure 1a) and magnetized (see Figure 5)
disks. Perturbations of the interchange mode make density and
magnetic flux change in phase, and thus activate fast
magneto-acoustic waves to propagate in the radial direction. Since
the fast waves tend to disperse enhanced density, the
gravitational instability triggered by the interchange mode ought
to grow slower than that by the undular mode. This is the reason
for the difference between the two modes.

We saw from Figure 2 that the existence of magnetic fields boosts
the gravitational instability triggered by the undular mode. To
check whether the same boosting is possible for the interchange
mode, we calculate $-\Omega^2$ of the odd-parity solutions for
$\alpha$ = 0, 1, 2, and 3. As shown in Figure 6, the maximum
growth rate decreases with $\alpha$. This trend is opposite to the
case of the undular mode. With the interchange perturbations the
magnetic fields have no means for adding extra material to the
densest region of the disk; instead they act to disperse the
enhanced density by driving the fast magneto-acoustic waves.
Therefore, the growth rate of the interchange mode decreases with
increasing field strength.

In all the cases illustrated in Figures 5 and 6, the magnetic
field is too strong to satisfy the criterion, $\gamma < 1 -
\alpha$, for convection. Originally this criterion is known to
hold for disks under external gravity (Newcomb 1961; Parker 1967;
Zweibel \& Kulsrud 1975; Ass\'eo et al. 1978; Lachi\`eze-Rey et
al. 1978). As for the $\gamma_{\rm c}$ criterion for the classical
Parker instability, Lee (2002) showed that the same convection
criterion, $\gamma < 1 - \alpha$, is applicable to the self
gravity case.

Adopting a weakly magnetized ($\alpha$ = 0.1) disk to see the
convective motions and assigning small values to the ratio of
effective specific heats to meet the convection criterion, we
calculated the dispersion relation as a function of $\nu_x$. The
results are shown in Figure 7, where the solid and the dotted
lines are for the odd- and the even-parities, respectively, and on
each curve $\gamma$ is marked. The well-defined peak corresponds
to the Jeans instability; while the monotonically increasing part
is owing to the convective instability. The even-parity
perturbations cannot trigger the gravitational instability but
certainly can activate the convection. After passing a minimum the
growth rate first increases monotonically with radial wave number
and reaches a finite value asymptotically as the wave number
further increases: Arrows outside the frame indicate the limiting
values of the growth rate. Under external gravities the
interchange mode perturbations of the shortest wavelengths are to
grow most rapidly and deprive the undular mode of the chance to
form large-scale structures. Under the self gravity, however, the
maximum growth rate of the gravitational instability can be higher
than the limiting growth rate of the convective instability. The
self gravity may thus have a chance fully to suppress convective
motions. This is a most important attribute of the
self-gravitating disk. Since the external gravity exerts
acceleration of fixed strength only in the vertical direction,
there is no means of de-structuring the vertical sheets generated
by the convective motions. Once a substantial degree of density
enhancement is achieved, however, the ever increasing self gravity
can provide the system with a means for redirecting the vertical
motions towards the dense mass center.

Even if the convection can grow faster than the Jeans instability
in linear stage of the disk evolution, particularly in `soft'
medium of small $\gamma$'s, the convective motion won't prevent
the system from forming dense structures in the midplane for two
reasons: First, no buoyancy force is available in the midplane and
hence no convective motions can take place there. Second, the
acceleration due to self gravity increases with density
enhancement. Therefore, once the gravitational instability is
triggered by odd-parity perturbations, it will eventually dominate
the disk over the convection.

The eigen-solution of the odd-parity interchange mode was
constructed for the case with $\gamma$ = 0.68 and $\nu_{x}$ = 0.6
($\lambda_{x}$ = 10.5$\,H$), which corresponds the well-defined
growth rate peak in Figure 7. The same was done of the even-parity
interchange mode with $\nu_{x}$ = 3.0 ($\lambda_{x}$ = 2.1$\,H$),
which is a convection dominating case. Figures 8a and 8b compare
the two sets of eigen-solutions in the $xz$-plane. The left and
right panels illustrate density and velocity fields, respectively,
and the color changes from dark red to white as density increases.
The numbers marked to thin solid lines on the left are logarithms
of $\rho_{1}(x,z)/\rho_{\mathrm {o}}(z=0)$. The thick solid lines
on the right demarcate the disk-halo boundary, the arrows inside
the frames represent velocity vectors, and the horizontal arrow on
top of each frame is a measure of the sound speed.

A well-defined density clump in Figure 8a is obviously a result of
the converging motions driven by the Jeans instability. The
velocity pattern in Figure 8a closely resembles the same pattern
in Figure 4a: Converging motions are evident all over the
$xz$-plane in Figure 8a and the $yz$-plane in Figure 4a. Figure 8a
suggests that once the interchange mode perturbation generates a
slight density enhancement at a midplane point, towards which the
self gravity would collect material and form a density clump. On
the other hand, we don't expect from Figure 8b such clump to form
in the midplane: Motions are absent from the low altitude region.

\subsection{Mixed Mode}

We now calculate the growth rate as a function of $\nu_{x}$ and
$\nu_{y}$. The results are presented in a three dimensional
surface plot of $-\Omega^2$ over the wave number domain. As our
interest is primarily in forming large-scale structures in the
Galactic ISM disk, we fix the disk thickness at $\zeta_{a}$ = 5,
which is sufficiently larger than the critical value, 0.549 (see
\S 3.1). Figures 9a and 9b are for {\it even}-parity solutions
with ($\alpha$ = 1, $\gamma$ = 1) and ($\alpha$ = 0.1, $\gamma$ =
0.8), respectively. The surface plot in Figure 9a looks like a
long `tunnel', and we already met, in Figure 1b, the dispersion
curve outlining the tunnel `entrance' at $\nu_{x}$ = 0.
Cross-section of the tunnel cut perpendicularly to the
$\nu_{x}$-axis describes the dispersion relation of the Parker
instability triggered by {\it undular} portion of the mixed mode
perturbation. Please note that the tunnel ridge inclines slowly
with $\nu_{x}$. This is owing to the interchange part of the mixed
mode. The {\it pure} interchange mode (i.e., $\nu_{y}$ = 0)
doesn't trigger the magnetic Rayleigh-Taylor instability (cf.,
Chandrasekhar [1961] for this terminology), since this disk does
not satisfy the $\gamma < 1 - \alpha$ criterion for this
particular mode. But {\it interchange} part of the mixed
perturbation does trigger the convection everywhere in the
($\nu_{x}$, $\nu_{y}$) plane except at $\nu_{y}$ = 0, since
another condition $\gamma < 1 + \alpha$ is required for truly
mixed mode. It is the convection that makes the `ridge' incline
with increasing $\nu_{x}$. In the case shown in Figure 9b, the
magnetic Rayleigh-Taylor instability is activated by the {\it
pure} interchange mode, as the disk satisfies now the condition
$\gamma < 1 - \alpha$. This is why the tunnel wall of Figure 9b is
partially `open' on the $\nu_{y}$ = 0 side.

The three-dimensional plots shown in Figures 9c and 9d are for the
{\it odd}-parity solutions but with the same set of system
parameters as in Figures 9a and 9b, respectively. We may relate
the two peaks located on the $\nu_{x}$- and $\nu_{y}$- axes to the
Jeans and the Parker-Jeans instabilities, respectively. The peak
with $\nu_{x}$ = 0 is significantly higher than the one with
$\nu_{y}$ = 0, since the undular mode perturbation provides the
Parker instability with favorable field configurations to assist
the Jeans. On the other hand, the pure interchange mode may not
get any boosting from the Parker instability. Consequently, the
pure Jeans peak on the $\nu_{x}$-axis is always lower than the
Parker-Jeans on the $\nu_{y}$-axis. The long ridge along $\nu_{y}$
$\simeq$ 1 is of course owing to the convection-assisted Parker
instability; as for the case of even-parity solutions, the
convection makes the ridge height incline with $\nu_{x}$. But the
ridge reaches a finite level in the limit
$\nu_x$$\rightarrow$$\infty$.

When mixed mode perturbations of the odd-parity are given to the
self-gravitating, magnetized, gaseous disk, the Jeans, Parker and
convective instabilities may all develop in the disk, and their
relative importance depends on the ($\nu_{x}, \nu_{y}$)
combination. We have constructed, from Figure 9c, a plot of equal
growth-rate contours in the ($\nu_x, \nu_y$) plane and presented
the resulting contour map in Figure 10. The thick dashed-line is
for a special contour, whose growth rate is equal to the the
highest ridge level: Any perturbations with wave numbers falling
within the dashed-line boundary would grow faster than the ones
outside. Therefore, with such perturbations, a collaboration of
the Parker and the Jeans instabilities may well suppress the
convection and the disk will bear out large-scale structures.

In a recent two-dimensional MHD simulation, Kim \& Ostriker
$\,$(2002) followed nonlinear responses of a self-gravitating,
magnetized, differentially rotating, thin disk to an externally
imposed spiral potential, and demonstrated regularly spaced spurs
jotting out initially in an almost perpendicular direction to the
arm. Although nonlinear interaction of the Parker-Jeans
instability and the spiral shock wave is out of the scope of this
paper, it should be pointed out that the highest point of the
growth rate surface is located on the $\nu_{y}$-axis$\,$(see
Figure 10). This is always so, if the disk is thicker than
$\sim$0.549$\,H\,$(see Appendix C). The resulting structure is,
therefore, of a cylindrical shape with its long axis being aligned
perpendicularly to the direction of magnetic field lines or spiral
arm$\,$(Nagai et al. 1998; Lee \& Hong 1999; Chou et al. 2000). It
is then suggestive that the cylindrical structures of this study
will make the spurs in the context of differential rotation and
density wave. The odd-parity undular mode seems to have tailored
the regular spacing and initial alignment of the spurs.

\section{FORMATION OF LARGE ISM STRUCTURES}

A replacement of driving agent of the Parker instability from the
external to the self gravity opens up a new route, through which
the instability may work constructively with the Jeans
gravitational instability to form large-scale structures in the
Galactic ISM disk. In a qualitative sense, the undular mode
odd-parity solutions seem to have all the means of alleviating the
difficulties involved in the time scale, length scale and density
enhancement of the classical Parker instability. In this section
we will extend our qualitative understandings of the dispersion
relation to making quantitative estimates of the various scales.

In this study the Galactic ISM disk is characterized by midplane
mean density $n_{\rm o}$ of hydrogen nuclei, {\it rms} dispersion
$c_{\rm s}$ of ``cloudlet" velocities, ratio $\alpha$ of magnetic
to gas pressure, and thickness $z_{\rm a}$ (= $\zeta_{\rm a}\,H$)
of the disk. Unless $\tanh \zeta_{\rm a}$ is substantially less
than unity, details of the dynamics involved with the Parker and
the Jeans instabilities do not depend on the thickness, which
leaves $n_{\rm o}$, $c_{\rm s}$ and $\alpha$ as the system
parameters of principal importance. We will examine how these
three parameters control the minimum growth time of the
Parker-Jeans instability, its corresponding length scale, and the
mass associated with the length.

\subsection{A Comparison Standard of the ISM Parameters}

To calculate dynamically important scales of the Parker-Jeans
instability, we need to establish a `standard' set for the system
parameters. By taking averages of various observations,
Ferri\`{e}re (1995, 1998a, 1998b) presented space-averages of the
ISM parameters, which may serve us a starting point for setting up
the standard. According to his 1998 results, the midplane
equilibrium number density of hydrogens in the solar neighborhood
is $n_{\rm o}$ = 1.17 cm$^{-3}$, which includes all forms of
hydrogens residing in the four ISM phases and the molecular medium
as well. With an inclusion of helium (9$\,$\% of hydrogen by
number), the mean molecular weight of ISM per hydrogen nucleus
becomes $\mu_{\rm ISM}$ = 1.36, which makes $n_{\rm o}\,\mu_{\rm
ISM}$ = 1.59$\,$cm$^{-3}$. As a comparison standard for density we
will simply take $n_{\rm o}\,\mu_{\rm ISM}\simeq$ 2$\,$cm$^{-3}$
(cf., Holmberg \& Flynn 2000; Kulkarni \& Heiles 1987).

Excluding the molecular medium, Ferri\`{e}re (1998a) determined
the total thermal pressure of the four ISM phase materials to be
4.81$\times$10$^{-13}\,$ergs$\,$cm$^{-3}$. As the one-dimensional
turbulent velocity he took 4.5$\,$km$\,$s$^{-1}$ and
6.9$\,$km$\,$s$^{-1}$ for the molecular and the cold neutral
media, respectively, and estimated the total interstellar
turbulent pressure to be
6.31$\times$10$^{-13}\,$ergs$\,$cm$^{-3}$. The sum of thermal and
turbulent pressures becomes then
11.12$\times$10$^{-13}\,$ergs$\,$cm$^{-3}$. If his estimate of the
cosmic ray pressure 9.6$\times$10$^{-13}\,$erg$\,$cm$^{-3}$ is
included in the pressure count, the total amounts to
20.7$\times$10$^{-13}\,$ergs$\,$cm$^{-3}$. With the standard
density $n_{\rm o}\,\mu_{\rm ISM}\simeq$ 2$\,$cm$^{-3}$, this
formally requires the isothermal sound speed or the velocity
dispersion to be 7.9$\,$km$\,$s$^{-1}$. This seems to be an
over-estimate of reality, particularly for ISM near the central
plane. Owing to the neglect of cosmic rays in this study, we will
include only the thermal and the turbulent pressures in the
isothermal equation of state and take $c_{\rm s}\simeq$
5.8$\,$km$\,$s$^{-1}$ as a better choice for the comparison
standard.

For the magnetic pressure Ferri\`{e}re (1998a, 1998b) gave
10.3$\times$10$^{-13}\,$ergs$\,$cm$^{-3}$. In terms of magnetic
field strength, this is equivalent to 5.1$\,\mu$G, which is
significantly stronger than frequently quoted 3$\,\mu$G (cf.,
Spitzer 1978). If we accept all the pressure values of his
estimate, $\alpha$ becomes 0.50; if the cosmic ray pressure is not
counted, the ratio becomes 0.93. On the other hand, if we take
3$\,\mu$G for the field strength and count only the thermal and
the turbulent pressures, then the magnetic-to-gas pressure ratio
becomes 0.3. Because many uncertainties are involved in deducing
the cosmic ray and the magnetic pressures from the observed
distribution of synchrotron emissivity (Ferri\`{e}re 1995, 1998a)
and because a proper treatment of cosmic ray behaviors is beyond
the scope of current study (cf., Ryu et al. 2003), $\alpha$ may
not be fixed at a specific value at this stage. Instead we will
examine how the scales vary with $\alpha$, and chose ($n_{\rm
o}\,\mu_{\rm ISM}\simeq$ 2$\,$cm$^{-3}$, $c_{\rm s}\,\simeq$
5.8$\,$km$\,$s$^{-1}$, 0 $\le \alpha \le$ 1) as the comparison
standard.

\subsection{Time, Length and Mass Scales}

A simple comparison of the odd-parity undular mode solutions to
the even-parity ones will tell us how much change the constructive
interplay between Parker and Jeans instabilities can bring to the
growth time and the length scale. We have listed in Table 1 the
normalized maximum growth rates ($\Omega^{\rm ODD}_{\rm max}$,
$\Omega^{\rm EVN}_{\rm max}$) and their corresponding wave numbers
($\nu^{\rm ODD}_{y,\,{\rm max}}$, $\nu^{\rm EVN}_{y,\,{\rm max}}$)
for a selected set of $\alpha$ values. Superscripts ``ODD" and
``EVN" can be identified with the Parker-Jeans and the Parker
instabilities, respectively. The table was prepared with $\gamma$
= 1.

The ratio given in the fourth column demonstrates that a sizeable
reduction is made in the growth time by the self gravity. In
weakly magnetized disks the time scale is reduced to one third the
case of pure Parker instability. Since long wavelength
perturbations are prone to trigger the gravitational instability,
we expect the length scale of the Parker-Jeans to be longer than
that of the Parker instability. Indeed the ratio, $\lambda^{\rm
ODD}_{y,\,{\rm max}}/\lambda^{\rm EVN}_{y,\,{\rm max}}\,$, in the
last column shows that an about 30$\,$\% increase in the length
scale can be achieved by the Parker-Jeans instability. The
relative increase in the length scale varies little with
$\alpha\,$; while the enhancement in the growth rate changes
significantly with the magnetic field strength.

The minimum growth time, $\tau_{\rm GRW}= (H/c_{\rm s})|
\Omega_{\rm max}|^{-1}$, is given by
\begin{equation}
\tau_{\rm GRW} \simeq 3.8\times 10^{7}\,{\rm years}
\left[{1+\alpha\over n_{\rm o}\,\mu_{\rm ISM}\,\,{\rm
cm}^{3}}\right]^{1/2}{1\over|\Omega_{\rm max}(\alpha)|}\, .
\end{equation}
Because the disk scale height is directly proportional to $c_{\rm
s}$, the time scale doesn't depend on it explicitly. Substituting
the $\alpha$-dependent $\Omega_{y,\,\rm max}^{\rm ODD}$ in Table 1
for the maximum growth rate, we examined how the growth time would
change with $\alpha$. Results are shown by thick solid line of
Figure 11. The two lower curves in thin solid lines are to be
identified with specific cases of $n_{\rm o}\,\mu_{\rm ISM}$ =
2$\,$cm$^{-3}$ and 3$\,$cm$^{-3}$. Please note that the system
parameter $n_{\rm o}\,\mu_{\rm ISM}$ is absorbed in the ordinate
for the thick solid line. To read the growth time scales for the
latter two cases directly off the thin lines, the $n_{\rm
o}\,\mu_{\rm ISM}$ factor should be ignored.

In terms of the system parameters the scale height $H$ becomes
\begin{equation}
H \simeq 1.9\times10^{2}\,{\rm pc}\left [{1+\alpha\over n_{\rm
o}\,\mu_{\rm ISM}\,\,{\rm cm}^{3}}\right ]^{1/2}\left [{c_{\rm
s}\over 5\,\,{\rm km\,s}^{-1}}\right ],
\end{equation}
which serves as the fundamental length for our disk system. The
perturbation wavelength, $\Lambda = 2\pi H\nu_{\rm max}^{-1}$,
corresponding to the maximum growth rate is given by
\begin{equation}
\Lambda \simeq 1.2\, {\rm kpc}\left [{1+\alpha\over n_{\rm
o}\,\mu_{\rm ISM}\,\,{\rm cm}^{3}}\right ]^{1/2}\left [{c_{\rm
s}\over 5\,\,{\rm km\,s}^{-1}}\right ]{1\over \nu_{y,\,\rm
max}(\alpha)}\, .
\end{equation}
The observed mean separation of the large-scale structures can be
compared directly with $\Lambda$, if $\nu_{y,\,\rm max}$ is
associated with the odd-parity solution. In the case of
even-parity, however, the observations should be compared with
$\Lambda/2$. When $\nu_{y,\,\rm max}^{\rm ODD}$ in Table 1 is
substituted for $\nu_{y,\,\rm max}(\alpha)$ in the above equation,
the length scale varies with $\alpha$ as in Figure 12, where
$n_{\rm o}\mu_{\rm ISM}$ and $c_{\rm s}$ are absorbed in the
calculation of ordinates for the thick solid line. To read the
length scales for the two lower curves please ignore the $n_{\rm
o}\,\mu_{\rm ISM}$ factor.

For a mass scale we count total amount of gas material that the
initial equilibrium disk had within a rectangular column of
cross-section $\lambda_{y,\,{\rm max}}\times\,\lambda_{x,\,{\rm
max}}$ extending vertically from $-z_{\rm a}$ to $+z_{\rm a}$. The
wave number $\nu_{y,\,{\rm max}}$ of the maximum growth rate fixes
azimuthal extent of the column cross-section. The undulation of
field lines under self gravity introduces to the system a natural
length scale in the azimuthal direction. However, it doesn't fix
the radial extent, since the highest growth rate point in the
dispersion-curve surface is always located on the $\nu_y$-axis.
(see, for example, Figure 10.) Only scale we may call in for the
radial extent is then the critical wavelength of marginal
stability or the width of spiral arm. In our normalization scheme
the critical wave number $\nu_{\rm crt}$ becomes 1 or very close
to it (see Appendix C), which gives a radial extent of
$\lambda_{x,\,{\rm max}}\,\simeq\, 2\pi\,H$. Since the total
column density of halo-bound isothermal gas disk is 2$H\rho_{\rm
o}(0)\tanh\zeta_{\rm a}\,$, we may have the mass scale, ${\cal
M_{\rm crt}}\,\simeq\,2H n_{\rm o}\,\mu_{\rm ISM}\,m_{\rm H}\tanh
\zeta_{\rm a}\cdot 2\pi H \nu_{y, {\rm max}}^{-1}\cdot 2\pi H$, as
\begin{equation}
{\cal M}_{\rm crt} \simeq 1.4\times 10^{7}\, {\rm M}_{\odot}\,
\left [{c_{\rm s}\over 5\,{\rm km}\,{\rm s}^{-1}}\right ]^{3}
\left [{(1 + \alpha)^{3}\over n_{\rm o}\,\mu_{\rm ISM}\,{\rm
cm}^{3}}\right ]^{1/2}{1\over \nu_{y,\,{\rm max}}\,(\alpha)}\,.
\end{equation}
When $\nu_{y,\,\rm max}^{\rm ODD}$ in Table 1 is substituted for
$\nu_{y,\,\rm max}\,(\alpha)$ in the above equation, the mass
scale varies with $\alpha$ as in Figure 13a, where the $n_{\rm
o}\,\mu_{\rm ISM}$ factor is absorbed again in the ordinate. Since
this mass scale depends on the velocity dispersion most
sensitively, we have illustrated, in the figure, for the following
four cases specifically: $c_{\rm s}$ = 1, 2, 3, and 5
km$\,$s$^{-1}$.

Since the disk scale height is about 200$\,$pc, the critical
wavelength often turns out larger than the width of spiral arm,
which is probably about a half kpc. For such cases, the arm width,
$\Delta R$, should be taken for the radial extent of the
rectangular column. The resulting mass scale, ${\cal M}_{\rm
arm}$, now reads
\begin{equation}
{\cal M}_{\rm arm} \simeq 1.2 \times 10^{7}\, {\rm M}_{\odot}\,
\left [{c_{\rm s}\over 5\,{\rm km}\,{\rm s}^{-1}}\right ]^{2}
\,\left [{\Delta R\over {\rm kpc}}\right ]\,\left [{1 +
\alpha\over \nu_{y,\,{\rm max}}\,(\alpha)}\right ]\,.
\end{equation}
It is interesting to notice that once the arm width is fixed by
the global dynamics of rotating disk, ${\cal M}_{\rm arm}$ doesn't
depend on the midplane density explicitly. Substituting
$\nu_{y,\,\rm max}^{\rm ODD}$ for the wave number of the maximum
growth rate, we have calculated the mass scale as a function of
the magnetic-to-gas pressure ratio, and shown the results in
Figure 13b for the four selected cases of the velocity dispersion.
This figure was prepared with $\Delta R$ = 1$\,$kpc.

\subsection{GMCs and HI Superclouds}

The two most prominent types of large interstellar structures are
the GMCs $\,$(Dame et al. 1986; Blitz 1993) and the HI
superclouds$\,$(Elmegreen \& Elmegreen 1987; Efremov \& Sitinik
1988; Elmegreen 1994). Clouds of each type show substantial
spreads in mass, size, mean separation, internal velocity
dispersion, and density. Particularly in the velocity dispersion
the two types almost overlap with each other$\,$(Falgarone \&
Lequeux 1973; Larson 1981; Heiles \& Troland 2003). However, the
GMCs differentiate themselves as a distinct type from the HI
superclouds in most of the directly observable
properties$\,$(Alfaro et al. 1992). In Table 2 we have summarized
observations of GMCs and HI superclouds separately. For entries of
the GMCs we relied heavily on Blitz (1993), and the HI supercloud
properties are mostly from Elmegreen \& Elmegreen$\,$(1987). The
mean separation of the superclouds is known to depend on mass: The
heavier are the clouds, the wider becomes their mean separation.

It is not easy to have a relevant time scale for structure
formation directly from observations, but a number of estimates
can still be made. For example, time required to pass an arm will
serve a good criterion for the growth time, since gathering
activity for the structure formation should take place mostly
while the gas material stays in a spiral arm. As suggested by Kim
\& Ostriker$\,$(2002), one half of the arm-to-arm moving time can
be a reasonable measure for the arm-crossing period. In an
$m$-armed spiral galaxy it will take $2\pi R/[mR(\Omega -
\Omega_{\mathrm P})]$ for the gas material to complete a journey
along the galactocentric circle of radius $R$ from one arm to the
next. Here, $\Omega$ is the angular velocity of the galactic
rotation at galactocentric distance $R$ and $\Omega_{\mathrm{P}}$
is that of the spiral pattern. With an approximation
$\Omega_{\mathrm{P}} \simeq \Omega$/2 (Binney \& Tremaine 1987)
the arm-passing time simply becomes $4\pi/(m\Omega)$. Keeping in
mind the solar neighborhood in the Galaxy, we take
220$\,$km$\,$s$^{-1}$ for the rotation velocity at $R$=8.5$\,$kpc,
and to the number of arms we may assign four$\,$(cf. Vall\'{e}e
1995) to have
\begin{equation}
\tau_{\mathrm{APT}} \simeq 60 \times 10^{6}\,{\mathrm {years}}\,
\left ({4\over m}\right )\left [{{26\,{\mathrm {km}}\,{\mathrm
{s}}^{-1}\,{\mathrm {kpc}}^{-1}}\over\Omega}\right ].
\end{equation}

Larson$\,$(1994) estimated the GMC growth time from the gas
consumption rate in the Galaxy. The star formation rate is in the
order of 3 M$_{\odot}$ per year, and about 2$\,$\% of cloud mass
is known to be converted into stars$\,$(Myers  et al. 1986; Evans
\& Lada 1991). This means that 150$\,$M$_{\odot}$ of gas is being
condensed, in every year, into star-forming molecular clouds.
Since the total amount of gas in the Galaxy is about
5$\times$10$^{9}\,$M$_{\odot}$, mostly in the form of GMCs, the
time required to collect this much gas into the GMCs is about
33$\times$10$^{6}\,$ years. One may then take 30$\times$10$^{6}$
years as the formation time scale of the GMCs.

We now have three time-scale values: The arm-to-arm crossing
period, 1.2$\times$10$^{8}$ years, is thought be a somewhat
generous time-scale criterion for the formation of HI superclouds,
while the arm-passing time, 60$\times$10$^{6}$ years, is a rather
strict one for them. On the other hand, the gas consumption
period, 30$\times$10$^{6}$ years, will serve as a reasonable
time-scale criterion for the formation of GMCs.

As is demonstrated in Figures 11, 12 and 13, the ranges of time,
length and mass scales that are expected from the undular
odd-parity solutions compare well with the observational criteria
of the HI superclouds summarized in Table 2. To be specific, with
the system parameter set, $n_{\rm o}\,\mu_{\rm ISM}\simeq$
2$\,$cm$^{-3}$, $c_{\rm s}\,\simeq$ 5.8$\,$km$\,$s$^{-1}$ and
$\alpha\simeq$ 0.5, we have the growth time $\tau_{\rm GRW}\simeq$
4.3$\times$10$^{7}\,$years, the mean separation $\Lambda\simeq$
1.9$\,$kpc, and the mass scale ${\cal M}_{\rm arm}\simeq$
2.8$\times$10$^{7}\,$M$_{\odot}$. With the same set, the disk
scale height becomes $H\simeq$ 190$\,$pc, and the resulting
critical wavelength, 1.2$\,$kpc, extends in the radial direction
beyond the spiral arm. As illustrated in the figures, the choice
of $\alpha$ = 0.5 won't make a critical issue in bounding the
various scales; the scales depend least sensitively on $\alpha$,
unless it is much bigger than unity. From the comparison we
conclude that under canonical conditions of the Galactic ISM the
Parker-Jeans instability can drive the ISM disk to form
condensations of the HI supercloud scale.

To make the theoretical scales agree with the observational
criteria of the GMCs, one has to strain the system parameters to
extreme degrees. For example, a trial set of $n_{\rm o}\,\mu_{\rm
ISM}\simeq$ 50$\,$cm$^{-3}$, $c_{\rm s}\simeq$
1.2$\,$km$\,$s$^{-1}$ and $\alpha\simeq$ 1.0 gives us 11 million
years for the growth time, 510$\,$pc for the mean separation, and
${\cal M}_{\rm crt}\simeq$ 1.4$\times$10$^{5}\,$M$_{\odot}$ for
the mass scale of the GMCs. As done in this example, one could
devise a combination of the system parameters that would {\it
formally} satisfy the observational criteria of the GMCs. But the
choice $n_{\rm o}\,\mu_{\rm ISM}\simeq$ 50$\,$cm$^{-3}$ seems too
high to be the density of {\it initial} equilibrium configuration.
And the choice of this high value of density requires the velocity
dispersion of cloudlets too low. The undular odd-parity solutions
under the {\it self gravity} alone may not grow the GMC and the HI
supercloud scale structures simultaneously.

\section{DISCUSSION AND CONCLUSION}

Through the present study we come to realize that in previous
investigations role of the interchange mode was unduly stressed in
its tendency of disrupting large-scale ISM structures that the
undular mode might otherwise have formed. The unfair emphasis was
the price to pay for both adoption of the external gravity as and
neglect of the self gravity from driving agent of the Parker
instability. In this paper we have examined what modifications
could be made to the classical picture of the Parker instability
by assigning the driving task solely to the self gravity.

If the vertical acceleration rendered by the adopted driving agent
is a constant of perpendicular distance from the central plane, it
may act as a source of buoyancy force over the entire disk.
Consequently, under such unrealistic uniform external gravity,
convective motions are to be activated everywhere in the disk, and
the interchange mode will indeed dominate the magnetized gas disk
over the undular one. In reality, however, the vertical
acceleration due mostly to the Galactic stars increases linearly,
from zero at the midplane, with the perpendicular distance up to
about a half kpc, beyond which it remains more or less constant.
This type of external gravity can not be a buoyancy source in the
disk midplane at least, where GMCs are mostly found; while the
convection is pronouncedly active, under the realistic non-uniform
gravity, only in high altitude regions, where the vertical
acceleration reaches its maximum.

The convection wouldn't then be an obstacle for the classical
Parker instability to form, for example, GMCs in the midplane, if
the undular mode could collect enough material there before the
interchange mode develops chaotic sheets at high altitudes.
However, as long as the buoyancy force is available, the growth
rate of the convection increases within the limiting value as the
perturbation wavelength gets smaller and smaller. Therefore, even
if all the realistic features are taken into account for the
external gravity, the convection triggered by the interchange mode
would develop small-scale features well before the classical
Parker instability excited by the undular mode generates any dense
large-scale structures in the disk central plane. Magnetic valleys
eventually anchor some material towards the central plane, but the
material takes form of vertically oriented thin sheets rather than
of dense clumps, because the external gravity has only vertical
component of acceleration. As long as the driving task is assigned
to the external gravity, there is no means to divert the
convective motions from the vertical direction and to form clumps
instead of sheets.

The thick dashed-line in Figure 10 represents the growth rate
contour corresponding to the highest ridge level. Any
perturbations whose wave numbers are within the dashed-line
boundary would grow faster than the ones outside. Therefore, with
such perturbations, a collaboration of the Parker and Jeans
instabilities may well suppress the convection and the disk will
bear out large-scale structures.

Once the self gravity takes over the driving task from the
external gravity, the effect of the Jeans gravitational
instability due to the self gravity appears on the dispersion
curves. The maximum growth rate of the Jeans instability at the
large-wavelength perturbations can be be higher than the limiting
growth rate of the convection at the small-wavelength
perturbations, depending on the adiabatic index (cf., Figure 7).
This is not what we've seen from the external gravity case.
Therefore, for long-wavelength perturbations of odd-parity, the
maximum growth rate of the undular mode becomes even {\it faster}
than the highest growth rate value of the interchange mode (cf.,
Figure 10). The odd-parity undular mode thus opens up a route in
the magnetized disk, through which the Parker and Jeans
instabilities work jointly to form large-scale condensations in
the midplane before the interchange mode disperses them into thin
sheets.

In this way the Parker-Jeans instability is expected to remove
major difficulties in forming the large-scale ISM structures. To
check whether this expectation can be fulfilled by the odd-parity
undular mode, we have compared the resulting time, length and mass
scales with the corresponding observations of the HI superclouds
and GMCs.

As far as the time scale is concerned, the Parker-Jeans
instability  grows quickly enough to satisfy the observational
criteria for the HI superclouds and the GMCs as well (cf., Figure
11). The formation time scales inferred from various observations
don't impose any tight limits on the system parameters,
particularly on $\alpha$, and the comparison done in Figure 11
doesn't seem to differentiate the HI superclouds from the GMCs
either. It was a bit surprise to us. Because of the neglect of
cosmic rays, we thought, the time scale of the Parker-Jeans would
meet the observational criteria only barely. In a recent study Ryu
{et al.}$\,$(2003) treated cosmic rays with the
diffusion-convection equation and realistic diffusion coefficients
for both directions, perpendicular and parallel to the field
lines. They have demonstrated that by pumping an extra buoyancy
into the disk system, cosmic rays make an about factor two
increase in the growth rate of the classical pure Parker mode,
{\it i.e.}, of the instability triggered by the undular
perturbation under the uniform external gravity. The factor two
may not be fully applied to the Parker-Jeans case, because the
uniform external gravity exaggerates buoyancy. Nevertheless, if
cosmic rays had been included, a substantial further reduction
would have been made in the growth time scale. It may worth
reminding that the Jeans gravitational instability was called in
to support the Parker in the first place. And the result was
successful, as was demonstrated by the $|\Omega_{\rm y,\,max}^{\rm
ODD}|/|\Omega_{\rm y,\,max}^{\rm EVN}|$ - ratios in Table 1: A
small increase in $|\Omega_{\rm y,\,max}^{\rm EVN}|$ will result
in a sizeable amplification through the ratio. An inclusion of
rotation wouldn't change the situation in any significant way,
because the rotation is known to decrease the growth rate by a few
to fifteen percents only$\,$(Shu 1974). Along this line of
reasonings, if 2$\,$cm$^{-3}$ is not an unrealistic choice for
$n_{\rm o}\mu_{\rm ISM}$ in the Galaxy, we believe that the time
scale we now have is an under-estimate for the Galactic ISM disk.
The reason for the underestimation seems to be in our neglect of
the external gravity.

As to the length scale, the Parker-Jeans instability seems to
operate more efficiently than what is required for the HI
superclouds. Under `canonical' conditions of the Galactic ISM with
$\alpha$ = 0.5, the odd-parity undular mode shows its maximum
growth rate at wavelength 1.5$\,$kpc, which is less than the
observed mean separation, 2.4$\,$kpc, of the HI
superclouds$\,$(McGee \& Milton 1964; Elmegreen \& Elmegreen 1983;
Alfaro et al. 1992; Efremov 1995). Because of the conceptual
difficulty in applying a single velocity dispersion to the entire
ISM disk, one may not consider this much difference in the length
scale to be serious. However, we think the difference is a
significant one, because existence of cosmic rays would reduce the
wavelength further. Therefore, comparison of the length scale also
suggests an over-subscription of the self gravity in our disk
system.

From dispersion relations shown in Figures 1 and 2 we notice the
growth  rate peak for the even-parity undular mode doesn't survive
in the competition with the odd-parity: The growth rate peak of
the pure Parker mode is overwhelmed by the Parker-Jeans peak. This
holds true for wide ranges of disk thickness and magnetic-to-gas
pressure ratio. As was illustrated in Figure 3, the Parker peak
may survive the Parker-Jeans overwhelming only in such
unrealistically soft media as the ones having $\gamma\le$0.7, for
example. In the magnetized gas disks where only the self gravity
is operative to excite instabilities with the undular mode, it
doesn't seem to be possible to have two distinct growth-rate peaks
in the dispersion curve, and the present study indicates that the
overwhelming Parker-Jeans maximum is strongly tied to the
structures of the HI supercloud scale, not to the GMC scale ones.
As Elmegreen(1994) pointed out, the characteristic mass and mean
separation of the HI superclouds are natural outcomes of the
gravitational instability in the self gravity dominating disk
systems.

To relate both the GMC and HI supercloud structures to the single
Parker-Jeans peak, one may rely on two-stage formation scenarios:
The HI supercloud forms first and some time later the GMCs emerge
as fragment products of the supercloud. This idea was suggested in
the previous section reluctantly, because of the GMCs that are
residing outside the supercloud. Almost all of the HI superclouds
contain GMCs, but not all the GMCs belong to the HI superclouds
(Elmegreen \& Elmegreen 1987; Efremov \& Sitnik 1988). Instead
refining the two-stage idea, we should first pay attention to the
external gravity heretofore ignored. In fact the external gravity
renders vertical acceleration about 5 times stronger than the one
expected from the equilibrium ISM stratification in our Galaxy
(Bienaym\'e et al. 1987; Ferri\`ere 1998a; Boulares \& Cox 1990;
Lee et al. 2004). Of course in later stages the ever increasing
self gravity of ISM eventually override the fixed external
gravity, but in the initial stage the self gravity is definitely
over-subscribed in our model disk. A balanced subscription between
the external and self gravities for the task of instability
driving would reduce dominance of the Parker-Jeans peak on one
hand and enhance the Parker maximum on the other. In this way the
odd-parity undular mode may exhibit two distinctly separated
growth-rate peaks in the dispersion curve, with the one at smaller
wave number corresponding to HI superclouds and the other at
larger number to GMCs.

From these discussions we may conclude the paper as follows: The
present study has reconfirmed that the Parker instability
triggered by the undular mode under external gravity may not be a
route towards the formation of such large-scale structures as the
GMCs in the Galaxy (Kim et al. 2004). Even under the self gravity,
the even-parity undular mode solution can not accommodate
important observational properties of the large-scale structures;
however, the odd-parity undular mode is shown to have all the
potentials to remove major difficulties that have been confronting
the classical Parker instability as the formation mechanism of the
large-scale structures. The resulting time, length, and mass
scales of the Parker-Jeans instability are shown to be in better
agreement with the HI superclouds than with the GMCs. An inclusion
of the external gravity along with a proper treatment of the
cosmic rays would refine the agreement towards the desired
direction.

\vspace{5mm}

It is our pleasure to acknowledge illuminating discussions and
stimulating conversations with W.-T. Kim, B. G. Elmegreen, J. Kim,
D. Ryu, and J. Franco. The authors are grateful to B.-C. Koo and
T. Nakagawa for guiding us to some important literatures. We are
also thankful to J. Pyo for his assistance in the preparation of
some of the diagrams. While preparing this paper, SSH was
benefited greatly from Astronomical Information Research Group of
the KASI, Korea and also from Laboratory Infrared Astrophysics of
the ISAS/JAXA, Japan. This study was supported in part by KOSEF
through grant R14-2002-058-01003-0.

\newpage
\appendix

\title{APPENDIX}
\section{Coefficients of Perturbation Equations}
The coeffients of the two second order ODEs [Eqs. (24) and (25)]
are as follow:
\begin{equation}
{A_2}  \equiv  \left(\omt^2 - {{2\alpha} \over {1+\alpha}}
\kyt^2\right) \biggl\{{{2\alpha+\gamma}\over{1+\alpha}} \omt^2-
{{2\alpha\gamma} \over {(1+\alpha)}^2} \kyt^2\biggr\},
\end{equation}
\begin{equation}
{A_1}  \equiv  -2\enskip\Theta\enskip \left(\omt^2 - {{2\alpha}
\over {1+\alpha}} \kyt^2\right)
\biggl\{{{2\alpha+\gamma}\over{1+\alpha}} \omt^2- {{2\alpha\gamma}
\over {(1+\alpha)}^2} \kyt^2\biggr\},
\end{equation}
\begin{displaymath}
{A_0}  \equiv -2 (1- \Theta^2) \omt^4 +{{4}\over{1+\alpha}}
\left[\alpha (1-\Theta^2) \kyt^2-\{(1-\alpha-\gamma) \kxt^2
+(1+\alpha-\gamma) \kyt^2\}\Theta^2\right]\omt^2
\end{displaymath}
\begin{equation}
+\,{{8\alpha
(1+\alpha-\gamma)}\over{(1+\alpha)^2}}\,\Theta^2\kyt^2 k^2
+\left(\omt^2-{{2\alpha}\over{1+\alpha}}\kyt^2\right)
\biggl\{\omt^4-{{2\alpha+\gamma} \over {1+\alpha}} k^2\omt^2
+{{2\alpha\gamma}\over{(1+\alpha)}^2} \kyt^2 k^2\biggr\},
\end{equation}
\begin{equation}
{B_1} \equiv i\omt^3 \left( \omt^2 -{{2\alpha}\over{1+\alpha}}
\kyt^2\right),
\end{equation}
\begin{equation}
{B_0} \equiv -i\omt\left[\biggl\{2
{{1+\alpha-\gamma}\over{1+\alpha}} k^2 -4{{\alpha}\over{1+\alpha}}
\kxt^2\biggr\}\, \Theta\,\omt^2 - {{4\alpha (1 +
\alpha-\gamma)}\over{(1 + \alpha)^2}}\Theta \kyt^2k^2 \right],
\end{equation}
\begin{equation}
{C_2} \equiv i\omt\biggl\{\omt^4 - {{2\alpha +
\gamma}\over{1+\alpha}} k^2\omt^2
+{{2\alpha\gamma}\over{(1+\alpha)^2}}\kyt^2k^2\biggr\},
\end{equation}
\begin{equation}
{C_0} \equiv -k^2 i\omt \left[ \omt^4+\biggl\{2 (1-\Theta^2)
-{{2\alpha+\gamma}\over{1+\alpha}} k^2 \biggl\} \omt^2
+{{2\alpha}\over{1+\alpha}}\biggl\{ {{\gamma}\over{1+\alpha}} k^2
-2 (1-\Theta^2)\biggr\}\kyt^2 \right],
\end{equation}
\begin{equation}
{D_1} \equiv 2 (1-\Theta^2)\omt^2 \left(\omt^2
-{{2\alpha}\over{1+\alpha}} \kyt^2\right),
\end{equation}
and
\begin{equation}
{D_0} \equiv -4 \Theta (1-\Theta^2)
\biggl\{\omt^4+{{1-\alpha-\gamma}\over{1+\alpha}} k^2 \omt^2
-{{2\alpha (1+\alpha-\gamma)}\over{(1+\alpha)^2}} \kyt^2
k^2\biggr\}.
\end{equation}
Here $k^2$ means $\kxt^2$+$\kyt^2$, and the other symbols have the
same meanings as in \S$\,$2 of the main text.

\section{Stable Modes}
In self-gravitating, magnetized, gaseous disks, we expect to have
such stable wave modes as gravity wave, Alfv\'en wave, and fast
and slow MHD waves. To check whether our system of the linearized
perturbation equations indeed has all these modes, we take the
equations in various limits of perturbation wave numbers.

\subsection{Vertical Gravity Wave Mode}
The vertical gravity mode can be isolated by taking the
perturbation equations in the limit $\nu_{x}$ = 0 and $\nu_{y}$ =
0. This wave travels vertically and is related to convective
motions$\,$(Lamb 1945).

We eliminate $\psi$ from equations (24) and (25) and put $\nu_{x}$
= 0 and $\nu_{y}$ = 0 in the resulting equation for $u_{z}$ to
have
\begin{equation}
{d^3\over{d\zeta}^3}u_z-2{d\over{d\zeta}}\left[\Theta{d\over{d\zeta}}u_z\right]
-{{1+\alpha}\over{2\alpha+\gamma}}\Omega^2{d\over{d\zeta}}u_z=0.
\end{equation}
Since for odd-parity solutions
\begin{equation}
{d^2\over{d\zeta}^2}u_z\Big|_{\zeta=0}=0\hspace{5mm}
{\mathrm{and}}\hspace{5mm} u_z\Big|_{\zeta=0}=0
\end{equation}
hold true, we may integrate equation (B1) and have
\begin{equation}
{d^2\over{d\zeta}^2}u_z-2\Theta{d\over{d\zeta}}u_z
-{{1+\alpha}\over{2\alpha+\gamma}}\Omega^2 u_z=0.
\end{equation}
Changing variable $\zeta$ to $Z \equiv \tanh\zeta$, we rewrites
this as
\begin{equation}
(1-Z^2)^2{{d^2}\over{dZ^2}}u_z-4Z(1-Z^2){d\over{dZ}}u_z-\sigma^2
u_z=0,
\end{equation}
where $\sigma$ is defined as
\begin{equation}
\sigma^2 \equiv {{1+\alpha}\over{2\alpha+\gamma}}\Omega^2.
\end{equation}
We then substitute $D=(1-Z^2)^{-p}\,u_{z}(Z)$ for $u_{z}(Z)$ and
$\varpi = (1 + Z)$/2 for $Z$. With parameter $p$ defined by
\begin{equation}
\normalsize{\sigma^{2} = 4p\,(p-\half),}
\end{equation}
Equation (B4) reads
\begin{equation}
\varpi(1-\varpi){{d^2}\over{d\varpi}^2} D + \left\{2p + \half
-(4p+1)\varpi\right\} {d\over{d\varpi}}D - 4p^{2}D = 0\,,
\end{equation}
which is a hypergeometric differential equation. In terms of $Z$
its general solution is given by
\begin{eqnarray}
u_z(Z)&=&(1-Z^2)^{p}
\Big[C_{1}{\rf}\left\{2p,2p;2p+\half;\half(1+Z)
\right\}\nonumber\\
&+&C_{2}\left\{\half(1+Z)\right\}^{\half-2 p}
{F}\left\{\half,\half;\thalf-2p; \half(1+Z)\right\}\Big]\,,
\end{eqnarray}
where $\rf$ is a {hypergeometric function}, and $C_{1}$ and
$C_{2}$ are integral constants (Polyanin \& Zaitsev 1995).

From the boundary conditions (Eq. [B2]), we construct a set of two
simultaneous indefinite equations of $C_{1}$ and $C_{2}$. To have
nontrivial solutions for the integral constants, matrix of the
simultaneous equations should have vanishing determinant $G(p\,)$.
This condition gives us secular equation $G(p\,)$ = 0 with the
following definition of $G(p\,)$:
\begin{eqnarray}
{G(p)}&=&{{p^2(2p+1)^2}\over{(2p+\half)(2p+\thalf)}}
\,\rf\big(\half,\half;\thalf-2p;\half\big)
\,\rf\big(2p+2,2p+2;2p+\fhalf;\half\big)\nonumber\\
&-&\Big[(4p^2-\quota)\,\rf\big(\half,\half;\thalf-2 p;\half\big)
+{{2p-\half}\over{2(4p-3)}}
\,\rf\big(\thalf,\thalf;\fhalf-2p;\half\big)\hspace{8mm} \\
 &+&{{9}\over{8(4p-3)(4p-5)}}\,\rf\big(\fhalf,\fhalf;\shalf-2
p;\half\big)\Big] \,\rf\big(2p,2p;2p + \half;\half\big)\nonumber.
\end{eqnarray}

In Figure A1 we have plotted $G(p\,)$ against $p$. As can be seen
from the figure, $G(p\,)$ becomes zero at negative half integers
and also at zero. In terms of $p$ the resulting eigen-frequencies
$\Omega$ are given as
\begin{equation}
\Omega^{2} = {{\,2\alpha + \gamma}\over{1 +
\alpha}}\,\,2p\,(2p-1),
\end{equation}
where $p$ = $-\,3$$k\,(2k-1)/2$ with $k$ = 0, 1, 2, $\cdots$.

This mode of stable waves represents gravity waves propagating
vertically in the equilibrium disk ( cf., Eq. [50] in Shu 1974).
In the absence of magnetic fields Lamb$\,$(1945)  called this as
the atmospheric wave. This mode appears at the very beginning
stage of nonlinear MHD simulations.

\subsection{Fast MHD Wave}
For the limiting case $\xi$ $\ne$ 0 and $\eta$ = 0, perturbations
in the $x-$direction disturb density, pressure and magnetic fields
in such a way that the field strength and density show alternating
regions of compression and rarefaction. This is a signature of
simple acoustic waves. However, it is not easy to isolate this
type of acoustic waves near the dense midplane, where the
self-gravitational force is likely to override the restoring force
of the magnetic pressure. In order to isolate the mode clearly, we
apply the equation to the limit $z$$\rightarrow$$\infty$, where
the gravity force becomes very weak.

In this limit ($\xi$ $\ne$ 0, $\eta$ = 0, $z$ $\rightarrow$
$\infty$) one may easily isolate the fast MHD wave mode,
\begin{equation}
\Omega^2={{2\alpha+\gamma}\over{1+\alpha}}\xi^2,
\end{equation}
or
\begin{equation}
\omega^2=(v_{\mathrm A}^2+a_{\rm s}^2) k_x^2,
\end{equation}
where $a_{\rm s}^2$ means $\gamma$\,$c_{\rm s}^2$. The numerator
$2\alpha + \gamma$ of equation (B12) indicates a coupling of
Alfv\'en and sound waves. And the $\xi^{2}$-term suggests that the
wave propagates in the horizontal direction perpendicularly to the
initial magnetic fields. The denominator 1 + $\alpha$ comes from
the definition of the scale height, $H$.

\subsection{Alfv\'en and Slow MHD Waves}
To isolate the Alfv\'en and the slow MHD modes, it is better to
suppress the Jeans and the Parker instabilities. We thus push the
perturbation equations to the limit 1 $\ll$ $\eta$ $\ll$ $\xi$,
and have the following dispersion relation:
\begin{equation}
(2\alpha + \gamma)\,\Omega^4 - {{4\alpha(\alpha+\gamma)}\over{1 +
\alpha}}\,\eta^2 \Omega^2
+{{4\alpha^{2}\gamma}\over{(1+\alpha)^{2}}}\,\eta^{2} = 0.
\end{equation}
This can be decoupled into the Alfv\'en mode
\begin{equation}
\Omega^{2}={{2\alpha}\over{1 +\alpha}}\eta^{2}\,,\enspace\enspace
{\mathrm {or}}\enspace\enspace \omega^{2}=v_{\mathrm A}^{2}
{k_y}^{2}
\end{equation}
and the slow mode of the MHD waves
\begin{equation}
\Omega^{2} = {{2\alpha\gamma}\over{(1 + \alpha)(2\alpha +
\gamma)}}\,\eta^{2}\,, \enspace\enspace {\mathrm {or}}
\enspace\enspace \omega^{2} = {{v_{\mathrm A}^2
a_{s}^2}\over{v_{\mathrm A}^{2} + a_s^{2}}}\,{k_y}^{2}.
\end{equation}
The two waves propagate in the horizontal plane along the
direction of initial magnetic fields. The $2\alpha\gamma/(2\alpha
+ \gamma)$-term represents the cusp speed for the group velocity
of the slow wave.

Finally, the magnetized sound wave is derived from the
$y$-component of the momentum equation. The sound waves propagate
along the magnetic field direction, with the dispersion relation
given by
\begin{equation}
\Omega^{2}={{\gamma}\over{1 + \alpha}}\eta^{2}\,,\enspace\enspace
{\mathrm{or}}\enspace\enspace \omega^2={a_s}^{2} {k_y}^{2}.
\end{equation}

\section{Marginal Stability Analysis}
For thick disks the Parker instability may assist the Jeans
gravitational instability, so the maximum of the growth rate curve
is located on the $\nu_{y}$-axis in the plane of perturbation wave
numbers. This was demonstrated in Figures 9c, 9d and 10, where the
peak on the $\nu_{y}$-axis is higher than the one on the
$\nu_{x}$-axis. As the disk thickness decreases, the growth rate
peak on the $\nu_{y}$-axis decreases, since the Parker instability
is harder to get triggered in thinner disks. In the disk of the
critical thickness, the Parker instability disappears, and the two
peaks on the $\nu_{x}$- and $\nu_{y}$-axes become of the same
height. For the disk thinner than the critical thickness, the
magnetic fields obstruct the system from developing the Jeans
instability along the $y$-direction. So, the maximum of the growth
rate curve is now located on the $\nu_{x}$-axis.

As was shown in Figure 3, the dispersion relation of the
Jeans-Parker instability has two local maxima. The peak at smaller
wave number is for the Jeans instability, while the one at larger
number for the Parker instability. The dispersion curve crosses a
critical wave number, which separates the region of stability from
that of instability. The critical wave number comes into being due
to magnetic tension, and the Parker instability gets stabilized by
the tension. Thus, investigating the marginal state, we can find
the critical thickness for the disk that may trigger the Parker
instability.

With $\xi$ = 0 and $\Omega$ = 0, equation (21) describes the
marginal state of the Parker instability as
\begin{equation}
{{\mathrm d^2}\over{\mathrm d\zeta^2}}\,u_z - 2\Theta {{\mathrm
d}\over{\mathrm d\zeta}}\,u_z
+\left[{{2(1+\alpha-\gamma)(1+\alpha)}\over{\alpha\gamma}}\,\Theta^2-\eta^2\right]
u_z = 0,
\end{equation}
where all the symbols have their usual meanings. Applying equation
(C1) to the disk boundary and denoting the eigenvalue of the
equation by
\begin{equation}
EV =
{{2(1+\alpha-\gamma)(1+\alpha)}\over{\alpha\gamma}}\,\Theta_a^2 -
\eta^2,
\end{equation}
where $\Theta_{\rm a}$ = $\tanh \zeta_{\rm a}\,$, we have the
critical wavelength as

\begin{eqnarray}
\lambda_{\mathrm{crt}}  =  2\pi\,\,\left [ {(1 +
\alpha)\,c^{2}_{\mathrm {s}}\over {\,2\pi\, {\mathrm
{G}}\,\rho_{\mathrm{o}}(0)}}
                      \right ]^{1/2}
                            \, \left [ {{\,2(1 + \alpha -
\gamma)(1 + \alpha)}
                 \over{\alpha\gamma}}\,\Theta_{\rm a}^2 - EV \right ]^{-1/2}\,\,.
\end{eqnarray}
Because the critical wavelength, $\lambda_{\mathrm {crt, non}}$,
of the non-magnetized, isothermal, gaseous disk which is
infinitely extended in vertical direction is given by
\begin{equation}
\lambda_{\mathrm{crt, non}} = 2\pi\sqrt{{{c_s^2}\over{2\pi
G\rho_{\mathrm o}(0)}}}\,\,,
\end{equation}
we know, equation (C3) should reduce to Equation (C4), in the
limit $\alpha\rightarrow$ 0, $\gamma$ = 1 and $\Theta_{a}$ = 1.
This can be realized by simply assigning unity to $EV$. From this
we infer $EV$ = 1 is an eigenvalue of our halo-bound finite disk.
Therefore, we may have the normalized critical wave number of the
perturbation propagating along the magnetic fields as
\begin{equation}
\eta_{\mathrm{crt}} = \left [{{\,2(1 + \alpha - \gamma)(1 +
\alpha)}
                \over{\alpha\gamma}}\,\,\Theta_{a}^{\,2} - 1 \right ]^{1/2}.
\end{equation}
This nicely recovers the critical wave number, $\sqrt{2\alpha+1}$,
of a magnetized, isothermal, gaseous disk infinitely extended in
vertical direction$\,$(Stod\'olkiewicz 1963).

Since the Parker instability may not be triggered in a disk whose
critical wave number is zero, from equation (C5), we have the
critical adiabatic index for an onset of the Parker instability
as:
\begin{eqnarray}
\gamma_{\mathrm {crt}} = {{(1+\alpha)^{2}\,
\Theta_{a}^{\,2}}\,\over
           {\,\Theta^{\,2} + {\,1 \over\,\, 2}\,(1 + 2\,\Theta_{a}^{\,2})\,\alpha}}
\end{eqnarray}
which again recovers the well-known $\gamma_{\mathrm {crt}}$ =
$(1+\alpha)^{2}/(1+3\alpha/2)$ condition for the classical Parker
instability in an infinitely extended disk under external gravity
(Kim \& Hong 1998). In the disk with $\gamma < \gamma_{\mathrm
{crt}}$, we won't expect the Parker instability to take a place.
In Figure 3 we have illustrated how the material rigidity controls
the dispersion relation of the Parker-Jeans instability. The
dispersion curve usually shows two bumps, one from the Jeans and
the other the Parker. But the Parker bump becomes inconspicuous as
the adiabatic index increase, the even-parity dispersion curve in
dashed line almost coincides with the wave number axis when
$\gamma$ = 1.04, and completely disappears as $\gamma$ further
increases; from equation (C6) we indeed have $\gamma_{\mathrm
{crt}}$ = 1.05 with $\alpha$ = 0.1 and $\zeta_{\rm a}$ = 5.

In \S 3.1, we have noticed that the Parker instability may assist
the Jeans instability only in thick enough disks; thin disks may
not conceive the Parker instability in the first place. What is
then the critical thickness separating the thin from thick disk?
This can be answered from equation (C5) by finding out such
$\Theta_{a}$ (=$\tanh\zeta_{a}$) that may make $\eta_{\mathrm
{crt}}$ zero, namely,
\begin{equation}
\zeta_{\mathrm {a, crt}}=\tanh^{-1}\left [
                    {{\alpha\gamma}\over{2(1+\alpha-\gamma)(1+\alpha)}}
                             \right ]^{1/2}.
\end{equation}

When $\zeta_{a} < \zeta_{a, \mathrm {crt}}$, the Parker
instability cannot operate in the system, and the magnetic fields
instead obstruct the self-gravitational instability. Maximum of
the growth rate curve for the gravitational instability is
located, in this case, on the $\nu_x$-axis, so the gravitational
instability would generate a long cylinder whose axis is parallel
to the magnetic field lines. On the other hand, when $\zeta_{a}
> \zeta_{a, \mathrm{crt}}$, the Parker instability assists the
gravitational instability, and the Parker-Jeans instability is
operative in the system. Maximum of the growth rate curve for the
Jeans-Parker instability is placed now on the $\nu_y$-axis, so the
instability would align the cylinder axis perpendicularly to the
field lines. When $\zeta_{a}$ = $\zeta_{a, \mathrm{crt}}$, the
instability takes place, and the Parker instability cannot assist
the Jeans to grow faster along the fields. Thereby, the
perturbation propagating along the $y$-direction coordinates with
that propagating along the $x$-direction. The critical thickness,
when $\alpha$ = 1.0 and $\gamma$ = 1.0 for example, becomes 0.549
$H$. As shown in Figure A2, the undular mode ($\nu_{x}$ = 0)
dispersion curve in solid line looks almost the same as the
interchange mode ($\nu_{y}$ = 0) one in dashed line.

\clearpage

\begin{deluxetable}{cccccccccc}
 \tablecolumns{7}
 \tablewidth{0pt}
 \tablecaption{Comparison of the Odd- and Even-Parity Solutions}
 \tablehead{
 \colhead{$\alpha$} &
 \colhead{$|\Omega_{\mathrm{y,max}}^{\mathrm{ODD}}|$} &
 \colhead{$|\Omega_{\mathrm{y,max}}^{\mathrm{EVN}}|$} &
 \colhead{$\tau_{\mathrm{min}}^{\mathrm{ODD}}/\tau_{\mathrm{min}}^{\mathrm{EVN}}$}&
 \colhead{$\nu_{\mathrm{y,max}}^{\mathrm{ODD}}$} &
 \colhead{$\nu_{\mathrm{y,max}}^{\mathrm{EVN}}$} &
 \colhead{$\lambda_{\mathrm{y,max}}^{\mathrm{ODD}}$/$\lambda_{\mathrm{y,max}}^{\mathrm{EVN}}$}
  }
 \startdata
  0.00 & 0.665 & -     & -     & 0.476 &  -    &  -    \\
  0.25 & 0.722 & 0.259 & 0.359 & 0.547 & 0.750 & 1.371 \\
  0.50 & 0.766 & 0.372 & 0.486 & 0.621 & 0.838 & 1.350 \\
  1.00 & 0.835 & 0.513 & 0.614 & 0.749 & 1.000 & 1.334 \\
  2.00 & 0.924 & 0.675 & 0.730 & 0.943 & 1.239 & 1.313 \\
  3.00 & 0.981 & 0.765 & 0.780 & 1.093 & 1.417 & 1.296
 \enddata
\end{deluxetable}

\clearpage

\begin{deluxetable}{lcc}
 \tablecolumns{8}
 \tablewidth{0pt}
 \tablecaption{Observational Properties of GMCs and HI superclouds}
 \tablehead{
 \colhead{Properties} &
 \colhead{GMCs} &
 \colhead{HI Superclouds}
 }
 \startdata
  Mass [M$_\odot$] & (1$\sim$10)$\times$10$^5$\tablenotemark{a,\rm{b}} &  (1$\sim$40)$\times$10$^6$\tablenotemark{c}  \\
  Mean Separation [kpc] & 0.4 $\sim$ 0.6\tablenotemark{d} & $\sim$ 2.4\tablenotemark{e} \\
  Velocity Dispersion [km s$^{-1}$] & 0.4 $\sim$ 8.5\tablenotemark{f,\rm{g}} & 1.1 $\sim$ 8.1\tablenotemark{a} \\
  Mean Density [cm$^{-3}$] & $\sim$ 10$^2$ H$_2$\tablenotemark{d} & $\gtrsim$ 9 H\tablenotemark{c}\\
  Size [pc] & $\sim$ 45\tablenotemark{d} & $\sim$ 400\tablenotemark{c} \\
  formation time scale [years] & 30$\times$10$^6$ & 80$\times$10$^6$
  \enddata
  \tablenotetext{a}{ Larson 1981}
  \tablenotetext{b}{Dame et al. 1986}
  \tablenotetext{c}{Elmegreen \& Elmegreen 1987}
  \tablenotetext{d}{Blitz 1993}
  \tablenotetext{e}{Alfaro et al. 1992}
  \tablenotetext{f}{Heiles \& Troland 2003}
  \tablenotetext{g}{Falgarone \& Lequeux 1973}

\end{deluxetable}

\clearpage

\begin{figure}
\resizebox{\textwidth}{!}{\includegraphics{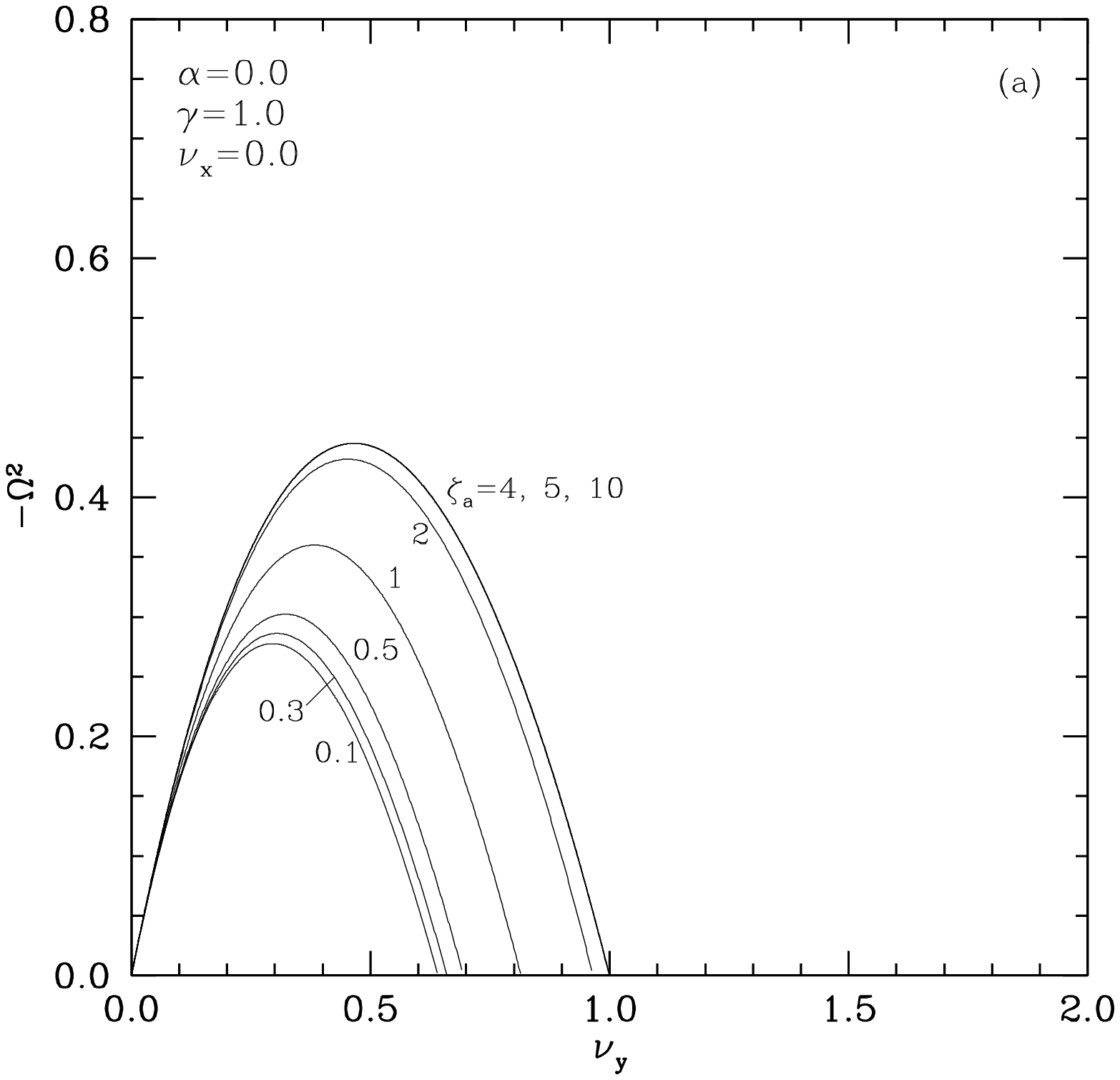}\includegraphics{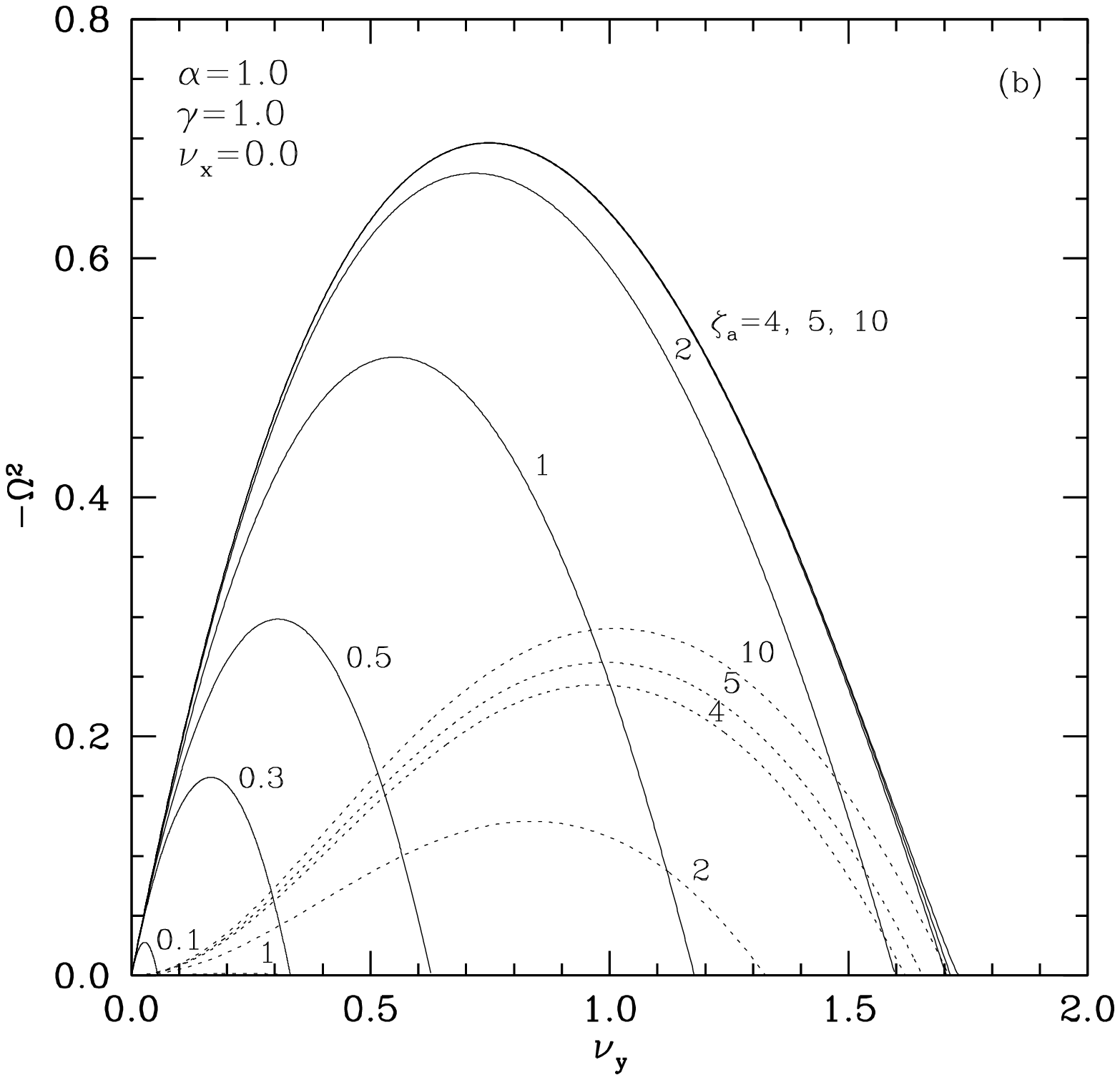}}
\caption{Dispersion relation of the undular mode ($\nu_{x}$ =
    0) for ({\it a}) non-magnetized
    and ({\it b}) magnetized gaseous discs of various thickness. The abscissa denotes
    effective wave number $\nu_{y}$, and the ordinate does square of the growth rate.
    Some of the disk and perturbation parameters are given in the upper left corner
    of each frame. The solid and dashed lines in ({\it b}) are for odd- and even-parity
    solutions, respectively. Because even-parity can not activate gravitational
    instability, there is marked suppression in the maximum growth rate from the odd-
    to the even-parity solutions.}
\end{figure}
\clearpage

\begin{figure}
\resizebox{\textwidth}{!}{\includegraphics{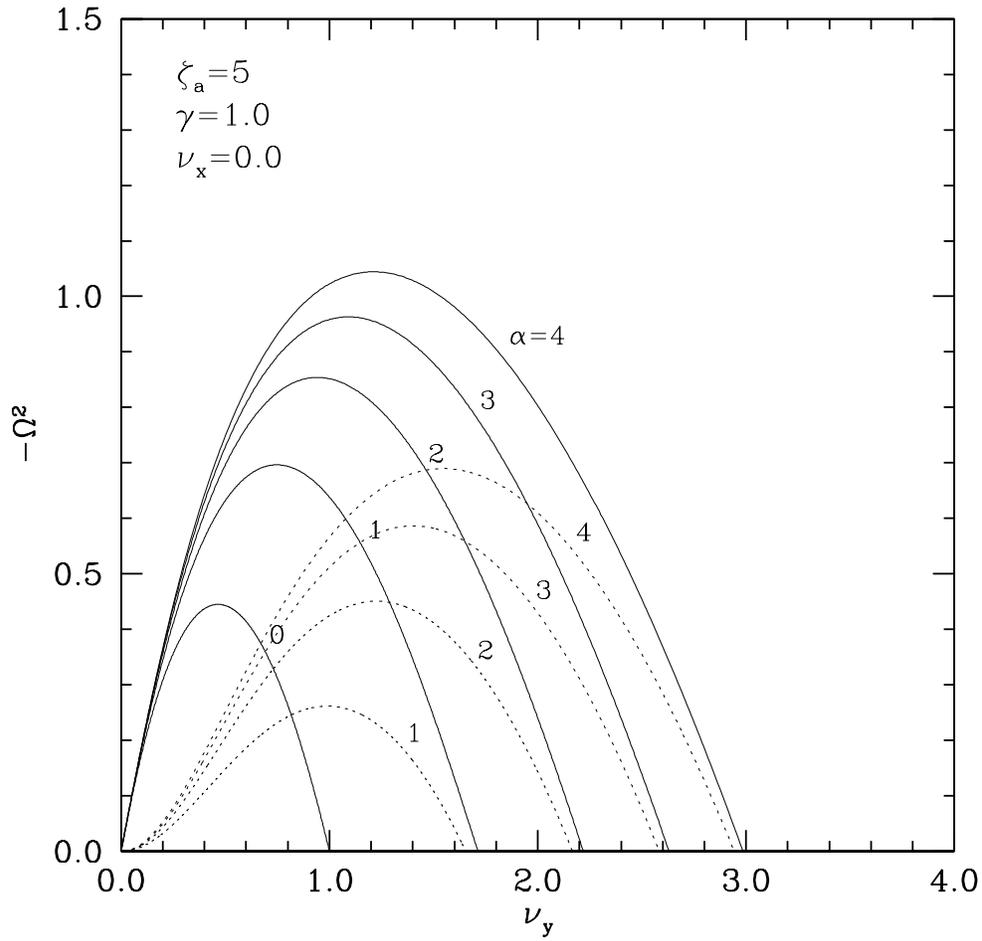}}
 \caption{Dispersion relation of the undular mode ($\nu_{x}$ = 0) for various values
    of $\alpha$. The abscissa denotes effective wave number $\nu_{y}$, and the ordinate
    does square of the growth rate. Some of the disk and perturbation parameters are given
    in the upper left corner. The solid and dashed lines are for odd- and even-parity solutions,
    respectively.}
\end{figure}
\clearpage

\begin{figure}
\resizebox{\textwidth}{!}{\includegraphics{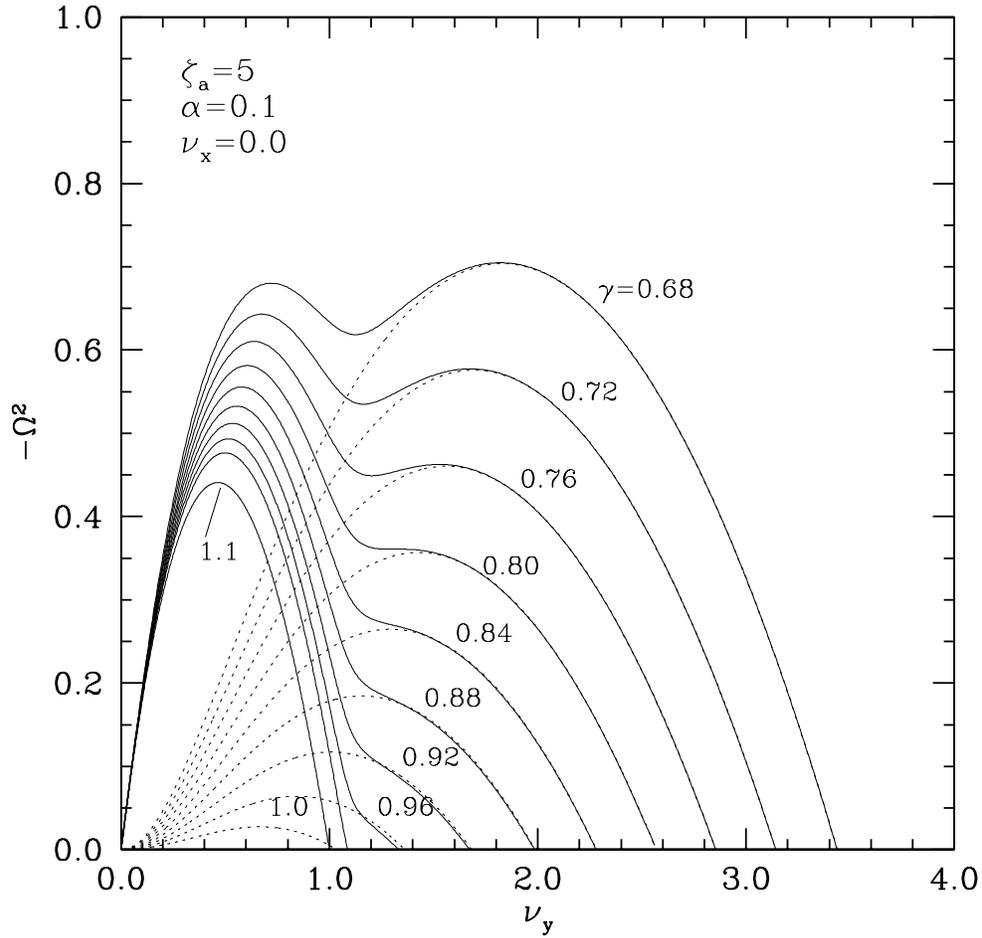}}
\caption{Dispersion relations for selected values of $\gamma$.
    The abscissa denotes effective wave number $\nu_{y}$ and the
    ordinate does square of the normalized growth rate of undular
    mode perturbation ($\nu_{x}$ = 0.0). Solid and dashed lines are
    for the odd- and even-parity solutions, respectively. The disk
    parameters are specified in the upper left corner, and $\gamma$-values
    are marked to each curve.}
\end{figure}
\clearpage

\begin{figure}
\epsscale{0.6} \plotone{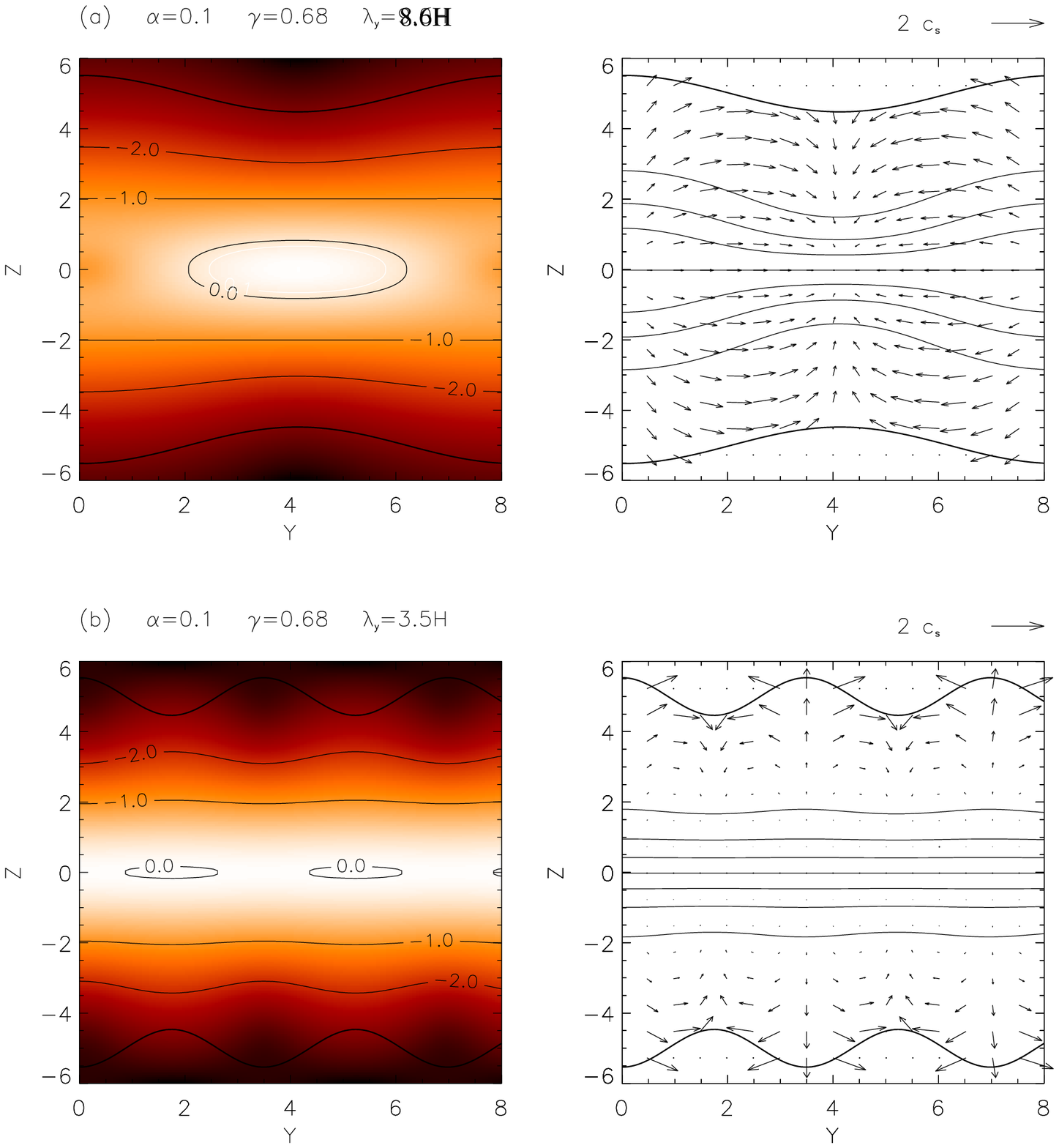}

\vspace{-5cm} \plotone{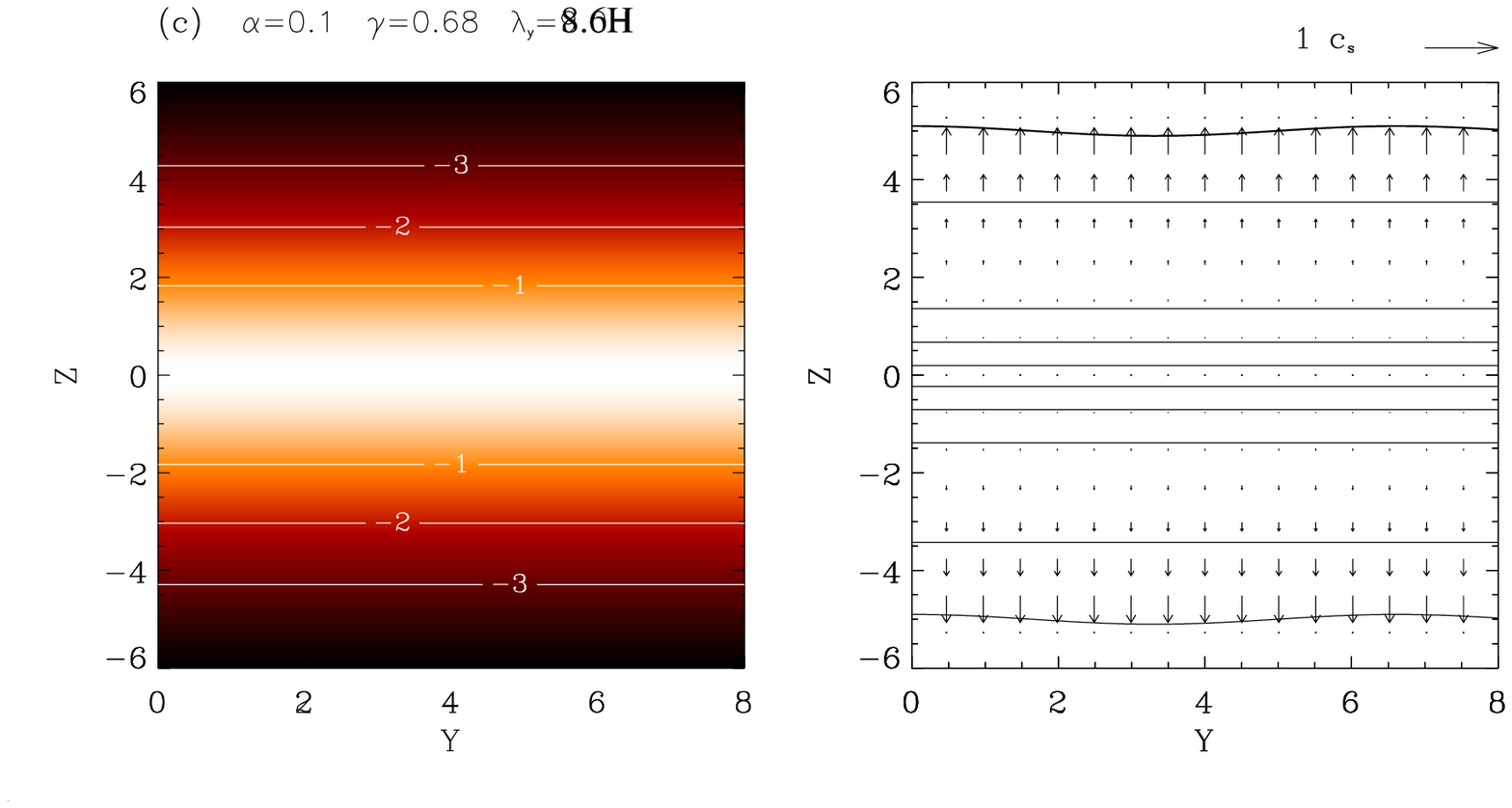}
 \caption{Eigen-solutions of the undular mode with odd-parity ({\it a},
    {\it b}) and with even-parity ({\it c}). Disk and perturbation parameters
    are specified on top of each panels on the left hand side. The abscissa and
    ordinate are $y$ and $z$ coordinates, respectively. The left hand side panels
    illustrate the density field, and color changes from dark red to white as
    logarithm of $\rho_{1}(y,z)/\rho_{\rm o}(z=0)$ increases. Arrow vectors
    in the right hand side panels illustrate the velocity field with magnitude
    reference being given on top of each frame. The thin solid lines represent
    the magnetic field lines and the thick lines do the disk-halo boundary. Case
    ({\it a}) is an example of the self-gravity dominating Parker-Jeans instability,
    while in ({\it b}) convection dominates, particularly in high altitudes. In
    case ({\it c}) the gravitational instability is not triggered, because
    even-parity condition is imposed upon the solution.}
\end{figure}
\clearpage

\begin{figure}
\resizebox{\textwidth}{!}{\includegraphics{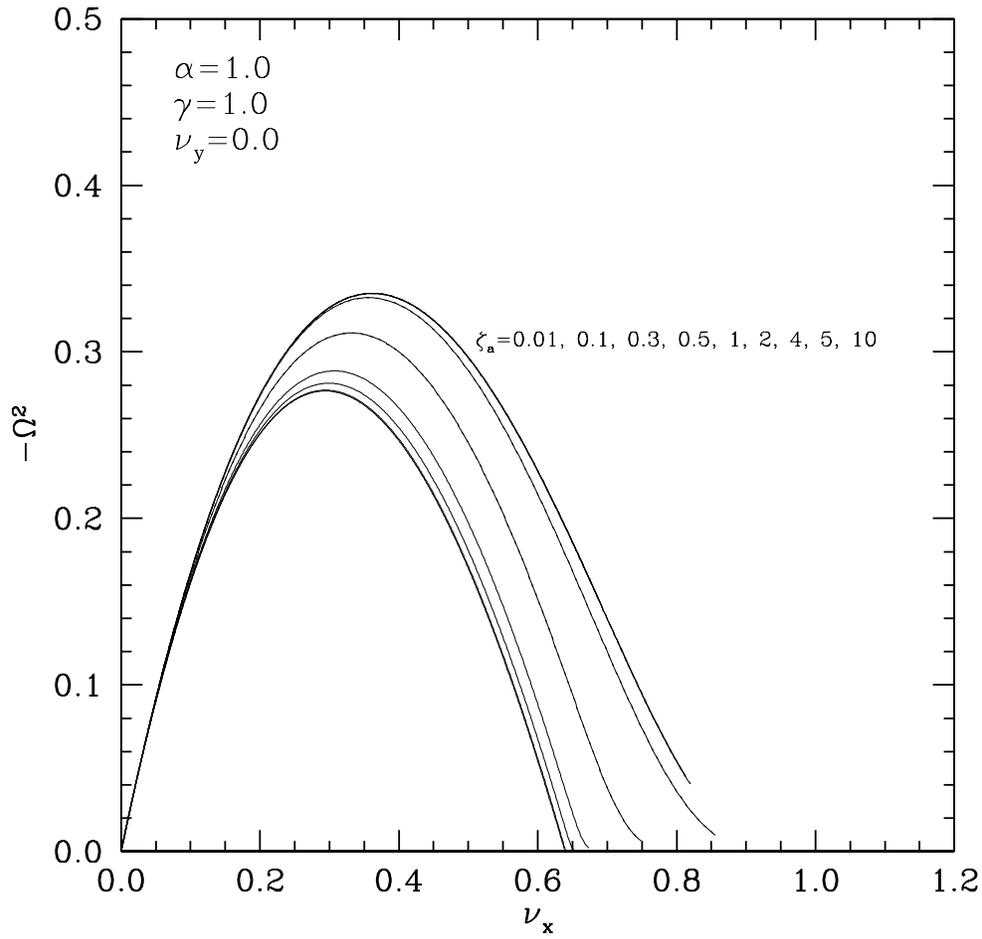}}
\caption{Dispersion relation of the interchange mode ($\nu_{y}$ =
0)
     for various disk thickness. The abscissa denotes effective wave
     number $\nu_{x}$ and the ordinate does the growth rate squared. To
     each curve is marked half thickness of the disk, and other disk
     parameters are given in the upper left corner.}
\end{figure}
\clearpage

\begin{figure}
\resizebox{\textwidth}{!}{\includegraphics{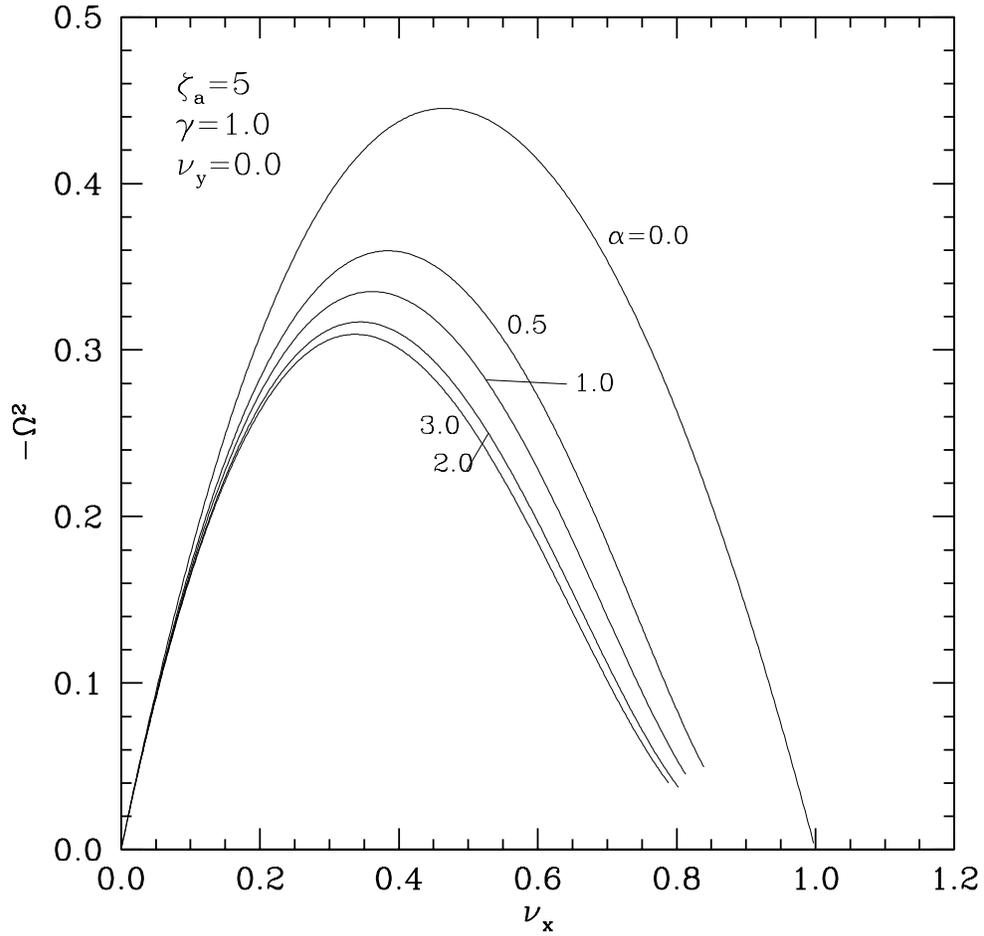}}
\caption{Dispersion relation of the interchange mode ($\nu_{y}$ =
0)
     for a number of selected $\alpha$'s. The abscissa denotes effective wave
     number $\nu_{x}$ and the ordinate does the growth rate
     squared. To each curve is marked $\alpha$ value, and other disk
     parameters are given in the upper left corner.}
\end{figure}
\clearpage

\begin{figure}
\resizebox{\textwidth}{!}{\includegraphics{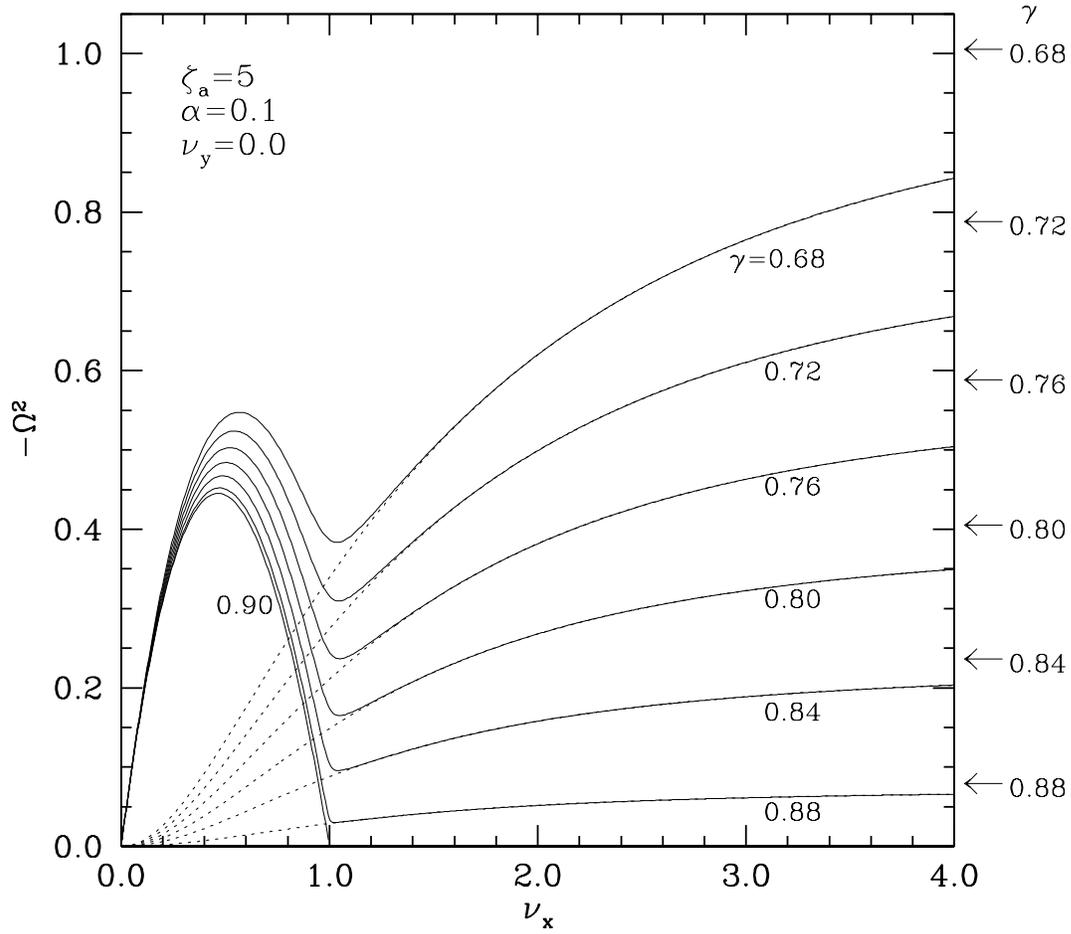}}
\caption{Dispersion relation of the interchange mode ($\nu$ = 0)
for selected $\gamma$
     values. The abscissa denotes effective wave number $\nu_{x}$ and the ordinate does the
     growth rate squared. To each curve $\gamma$ value is marked, and other disk parameters
     are specified in the upper left corner. The numbers with arrow on the right hand side
     of the frame indicate square of the maximum growth rate reached in the limit
     $\nu_{x}\rightarrow\infty$.}
\end{figure}
\clearpage

\begin{figure}
\resizebox{\textwidth}{!}{\includegraphics{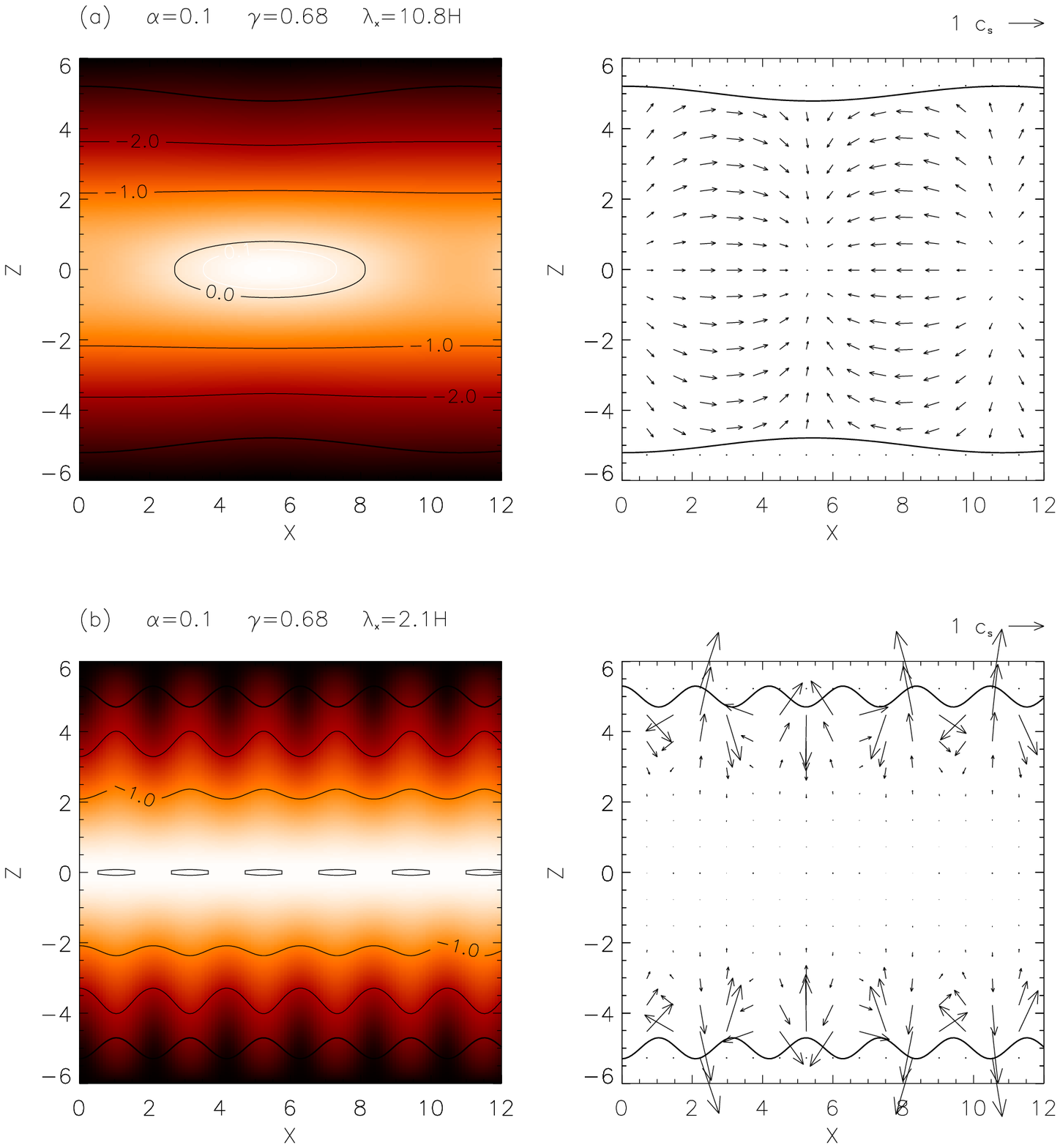}}
\caption{Eigen-solutions of the interchange mode with odd-parity
({\it
    a}) and with even-parity ({\it b}). Disk and perturbation parameters
    are specified on top of each panels on the left hand side. The abscissa and
    ordinate are $x$ and $z$ coordinates, respectively. The left hand side panels
    illustrate the density field, and color changes from dark red to white as
    logarithm of $\rho_{1}(y,z)/\rho_{\rm o}(z=0)$ increases. Arrow vectors
    in the right hand side panels illustrate the velocity field with magnitude
    reference being given on top of each frame. The thick lines represent the
    disk-halo boundary. Case ({\it a}) is an example of the Jeans gravitational
    instability triggered by long wavelength perturbation, while in ({\it b})
    the convective motions activated by short wavelength perturbation are
    apparent, particulary in high altitudes.}
\end{figure}
\clearpage

\begin{figure}
\resizebox{\textwidth}{7cm}{\includegraphics{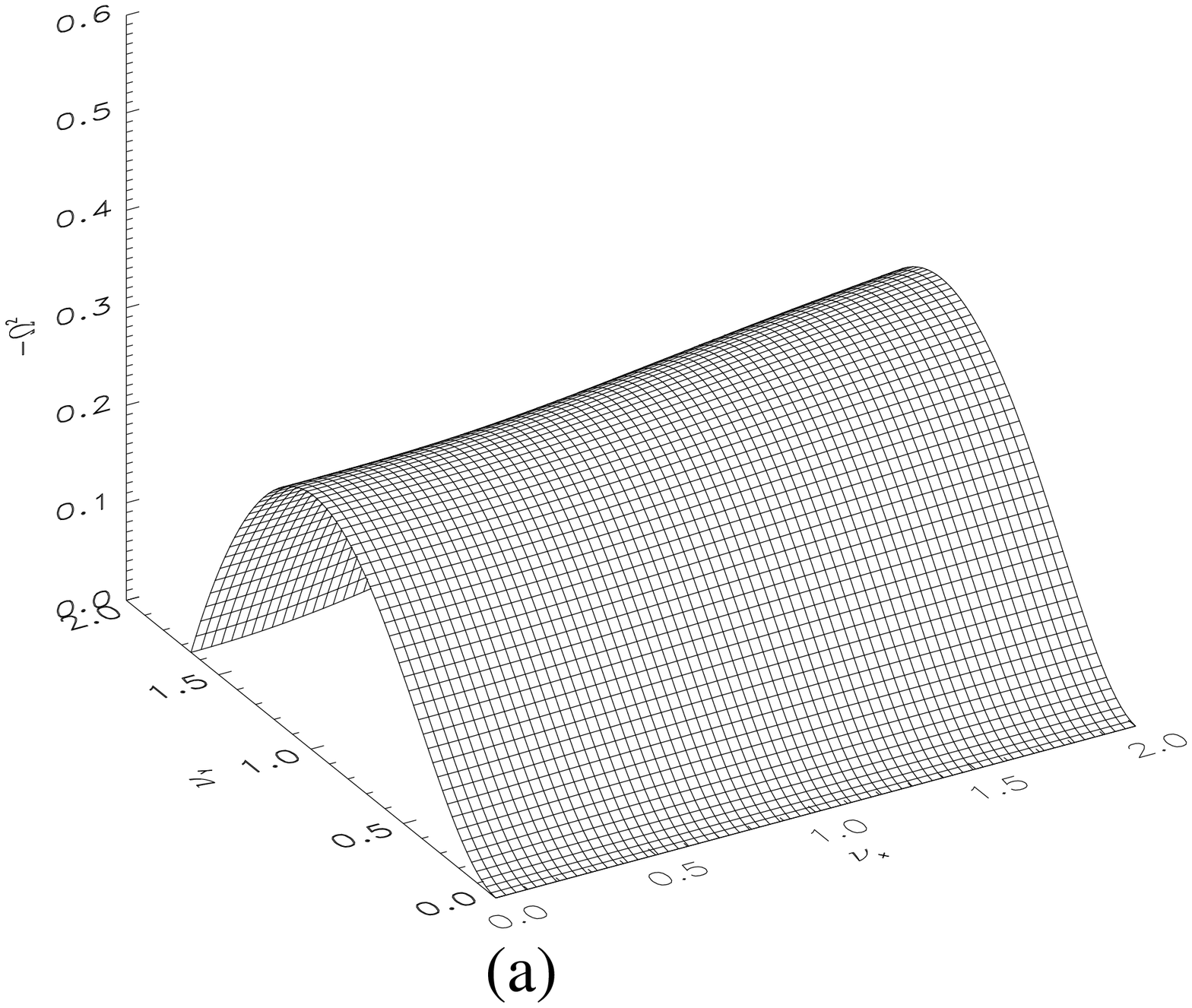}\includegraphics{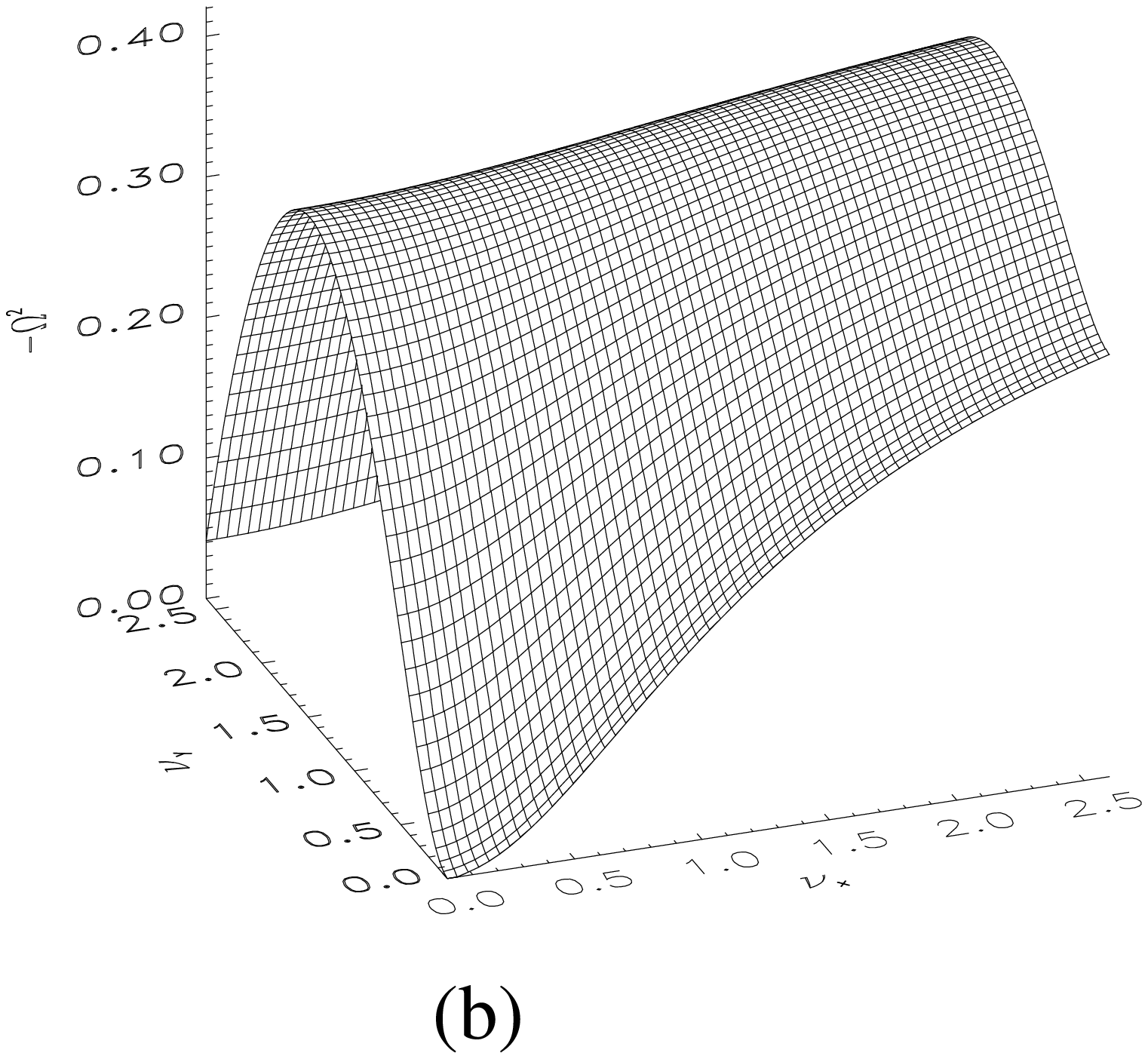}}
\resizebox{\textwidth}{7cm}{\includegraphics{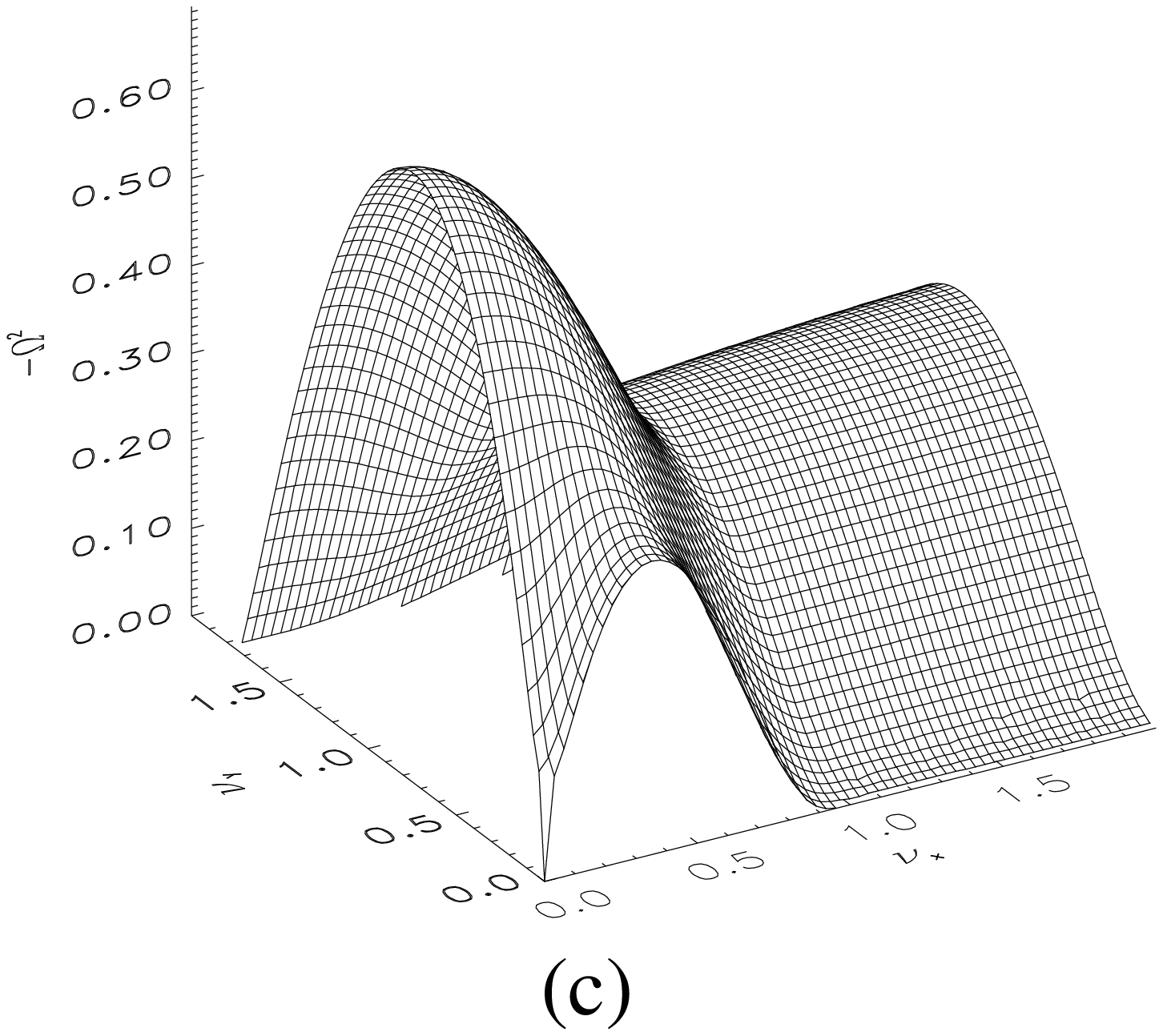}\includegraphics{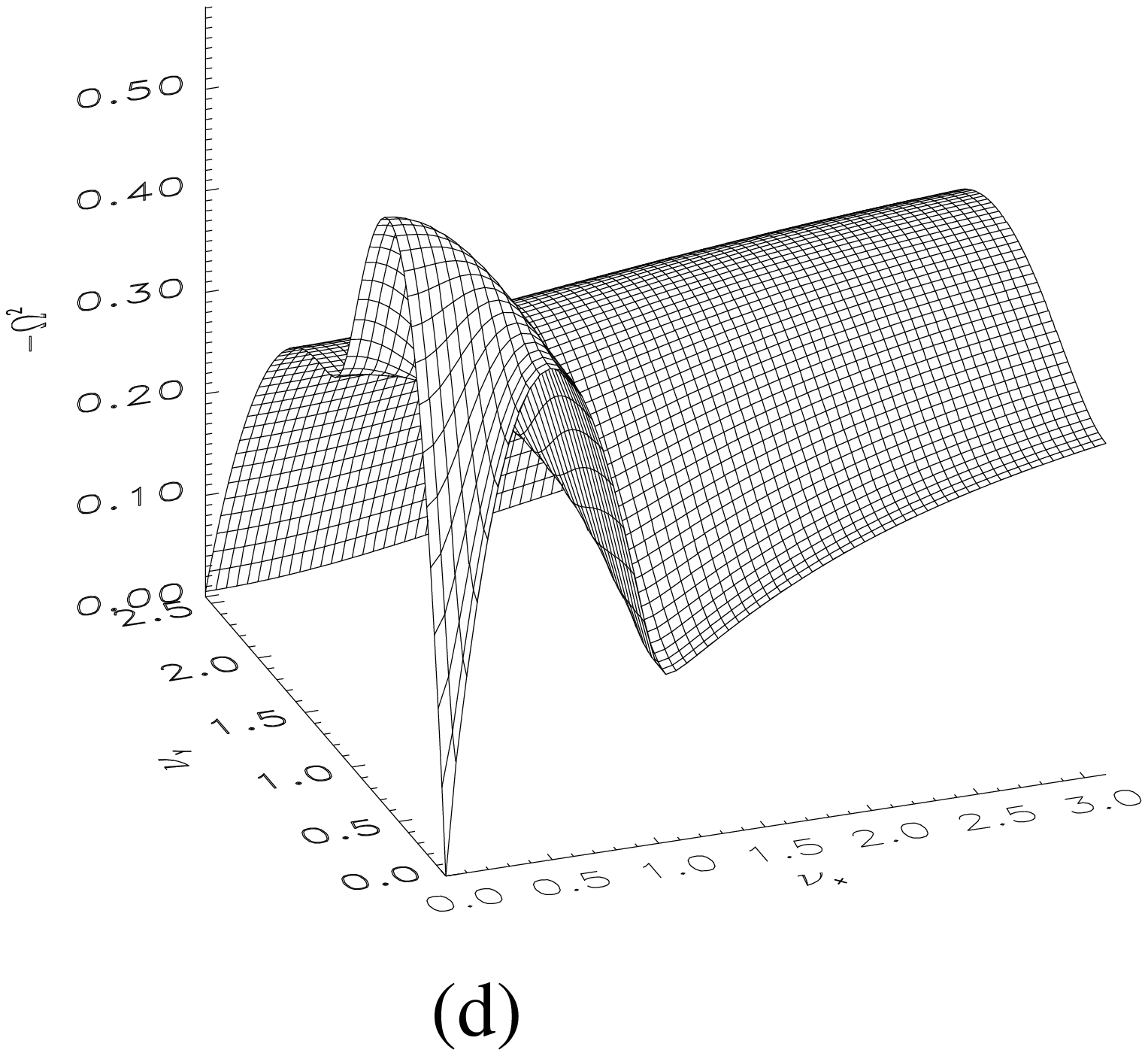}}
 \caption{A three-dimensional surface plot for the dispersion relation of the
     mixed mode. In all four cases the halo is placed at $\zeta_{a}$ = 5. Other
     parameters are as follow: In case ({\it a}) $\alpha$ = 1 and $\gamma$ = 1
     with even-parity; ({\it b}) $\alpha$ = 0.1 and $\gamma$= 0.8 with even-parity;
     ({\it c}) $\alpha$ = 1 and $\gamma$ = 1 with odd-parity; and ({\it d})
     $\alpha$ = 0.1 and $\gamma$ = 0.8 with odd-parity. The ordinate represents
     square of the growth rate and the two abscissae denote the horizontal
     perturbation wave numbers. In ({\it a}) and ({\it b}) the gravitational
     instability has not been activated; while in ({\it c}) and ({\it d}) the
     Parker-Jeans instability is operative. On the other hand, in ({\it b}) and
     ({\it d}), the convection criterion $\gamma < 1 - \alpha$ for the pure
     interchange mode is fulfilled, and hence the Rayleigh-Taylor instability
     gets triggered. The inclining `ridge' of the growth-rate surface reaches
     in the $\nu_{x}\rightarrow\infty$ a finite level, which is lower than the
     Parker-Jeans peak on the $\nu_{y}$ axis.}
\end{figure}
\clearpage

\begin{figure}
    \resizebox{\textwidth}{!}{\includegraphics{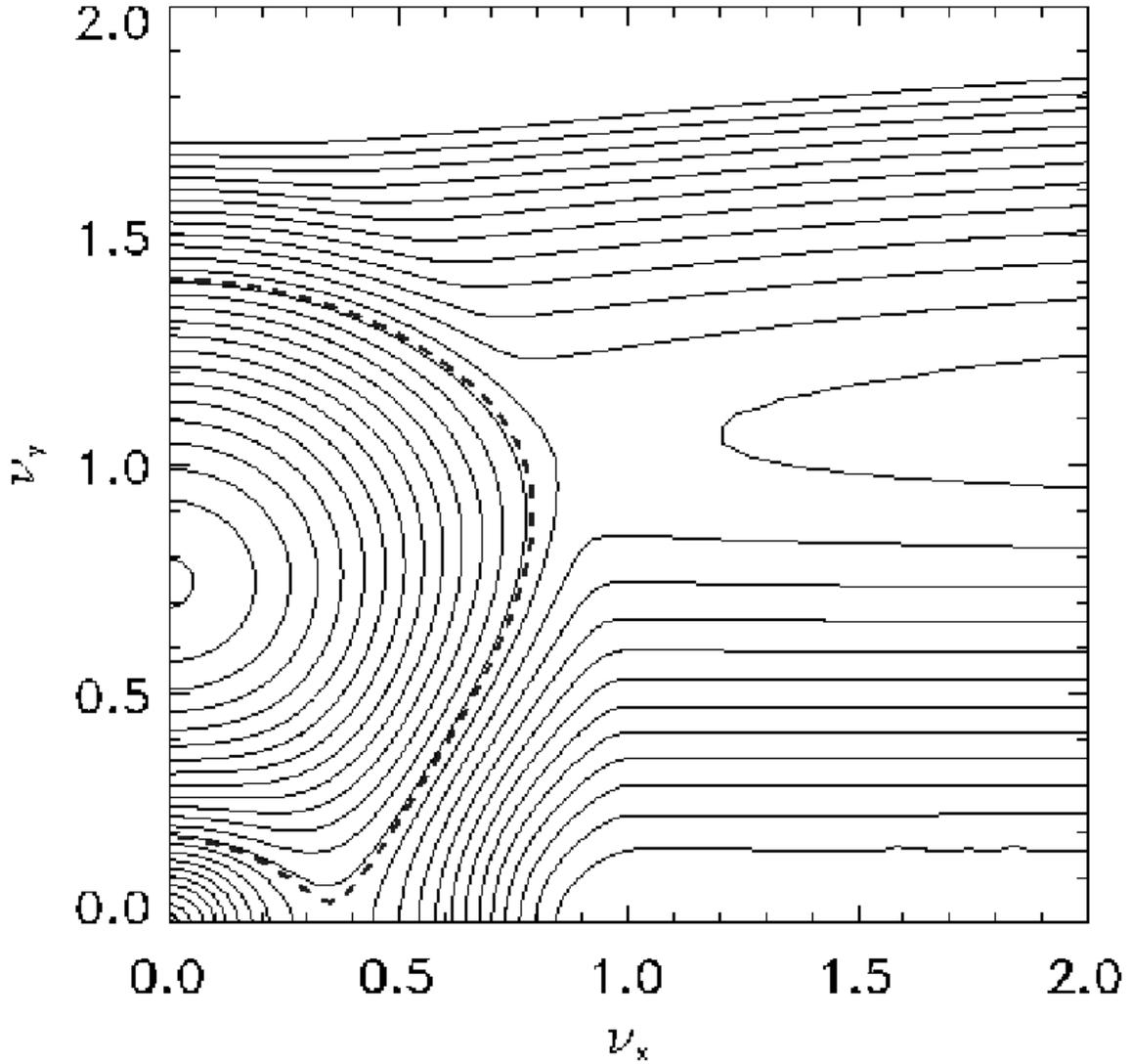}}

    \vspace{0cm}
    \caption{Equal growth rate contour map. This is constructed from the
    3-dimensional surface plot of the dispersion relation for undular
     odd-parity solutions with $\zeta_{\rm a}$ = 5.0, $\alpha$ = 1.0,
     $\gamma$ = 1.0. The dashed line contour in blue corresponds to the
     highest ridge level. Therefore, the perturbations whose wave number
    pair falls within the blue contour are expected to grow faster than the
   convective instability.}
\end{figure}
\clearpage

\begin{figure}
    \resizebox{\textwidth}{!}{\includegraphics{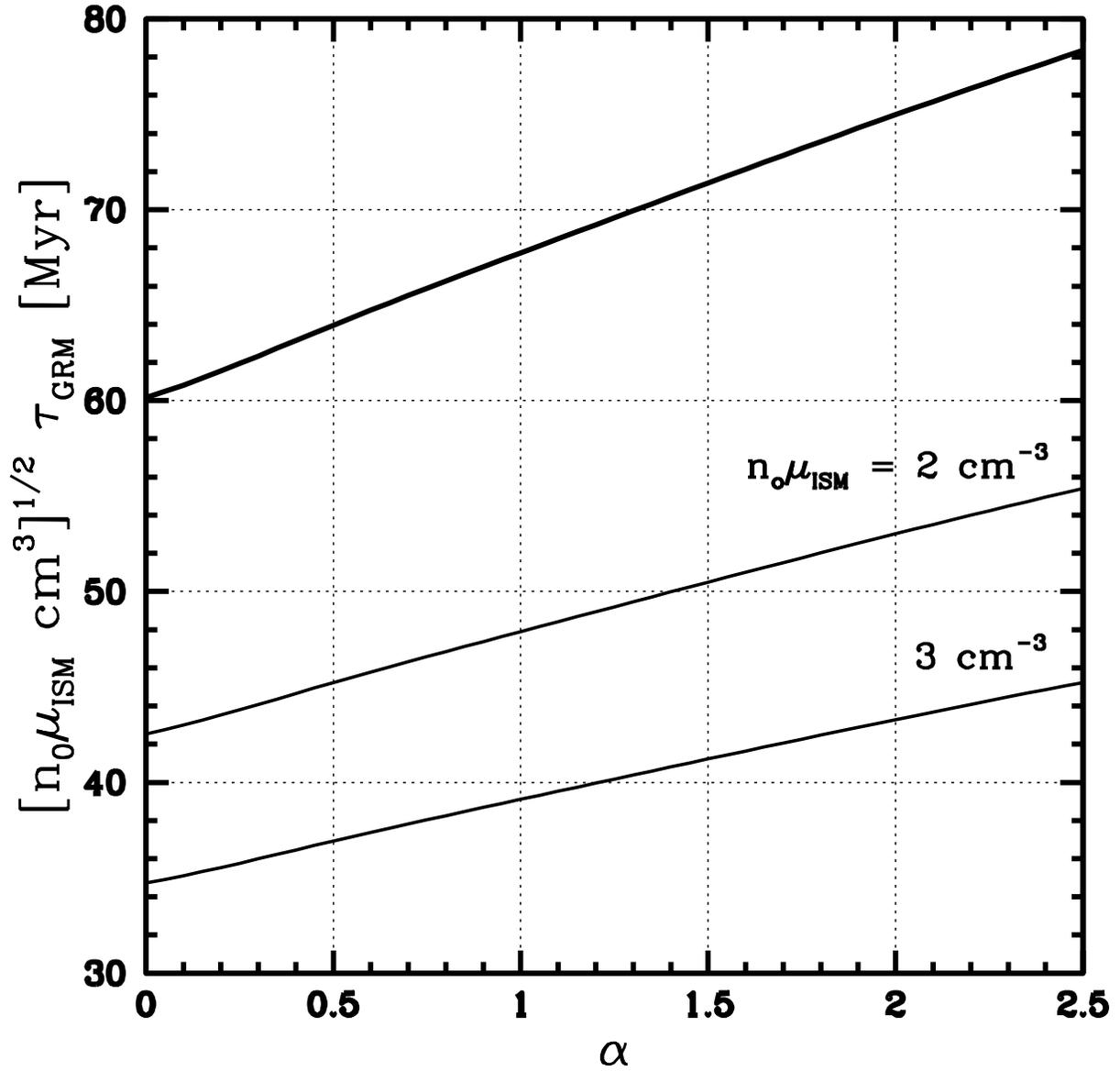}}
    \caption{Growth time scale {\it versus} magnetic-to-gas pressure ratio.
     Notice that the $n_{\rm o}\,\mu_{\rm ISM}$ factor is absorbed in the
     ordinate. When the ordinates for the lower two curves in thin solid line
     are read directly off the figure, the density factor should be ignored. }
\end{figure}
\clearpage

\begin{figure}
    \resizebox{\textwidth}{!}{\includegraphics{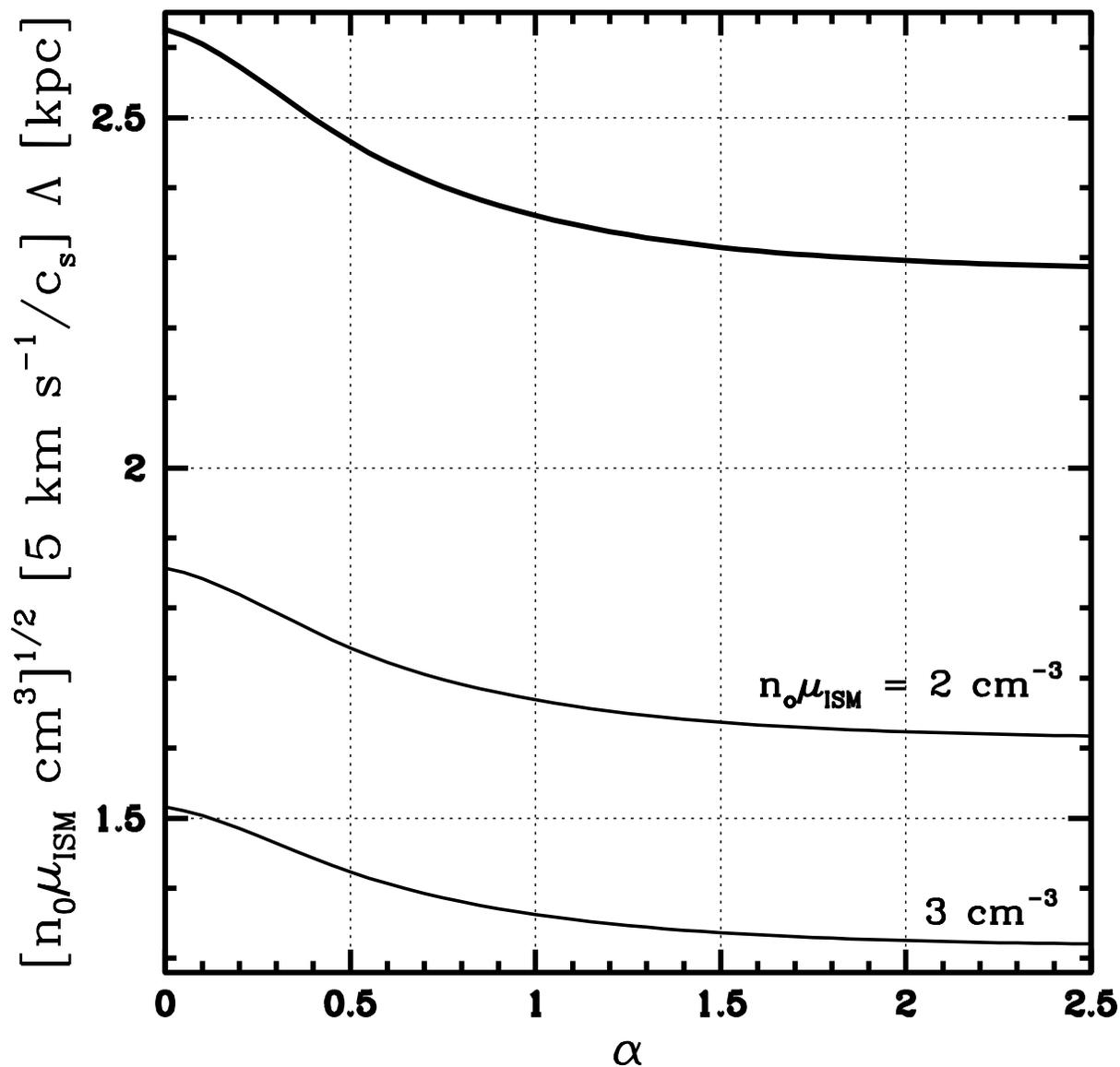}}
    \caption{Wavelength of maximum growth rate  {\it versus} magnetic-to-gas
     pressure ratio. Notice that $n_{\rm o}\,\mu_{\rm ISM}$ and velocity
     dispersion are absorbed in the ordinate for thick solid line. When the
     ordinates for the lower two curves in thin solid line are read, the
     multiple of the two factors should be considered unity.}
\end{figure}

\begin{figure}
    \resizebox{\textwidth}{!}{\includegraphics{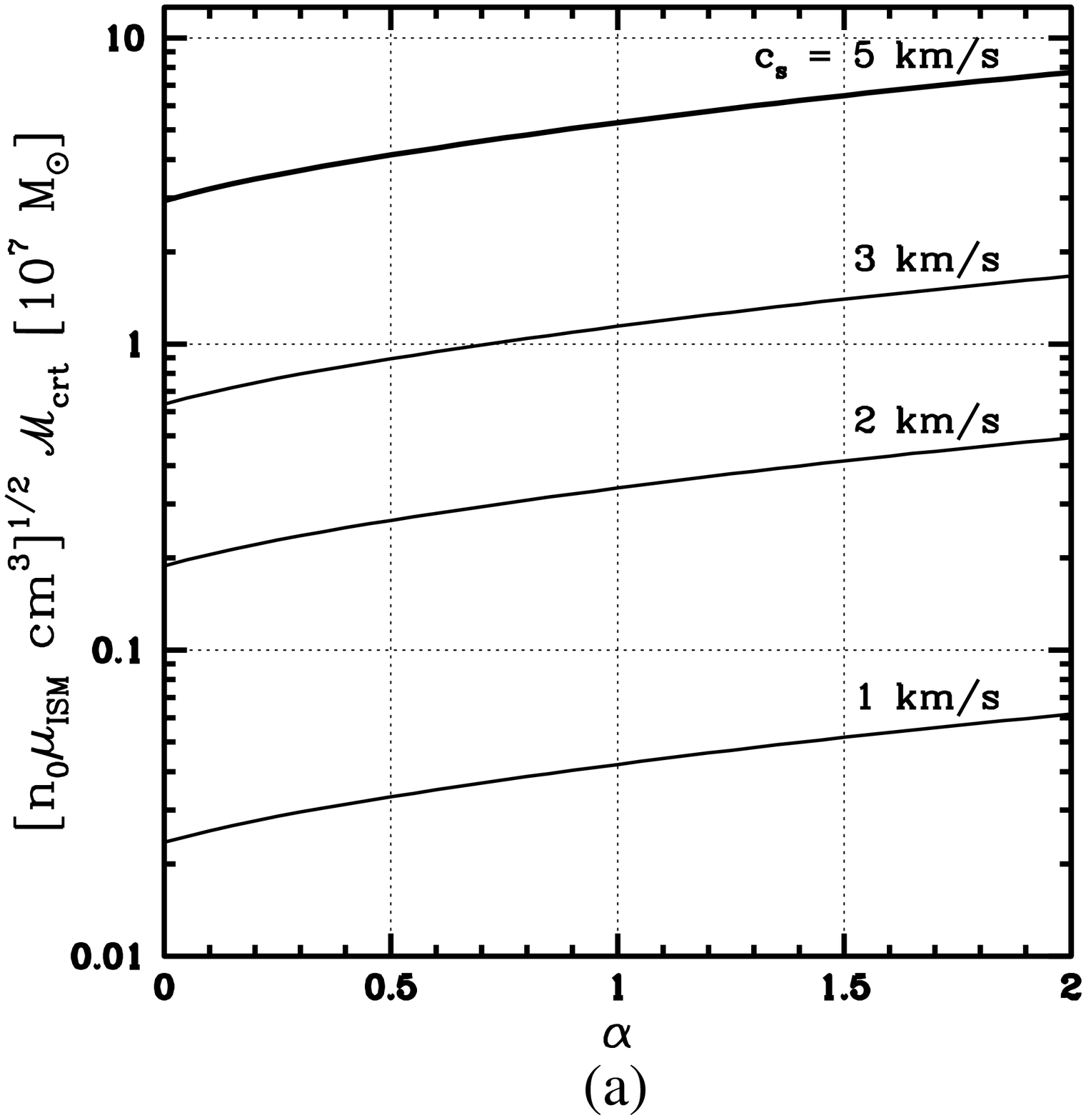}\includegraphics{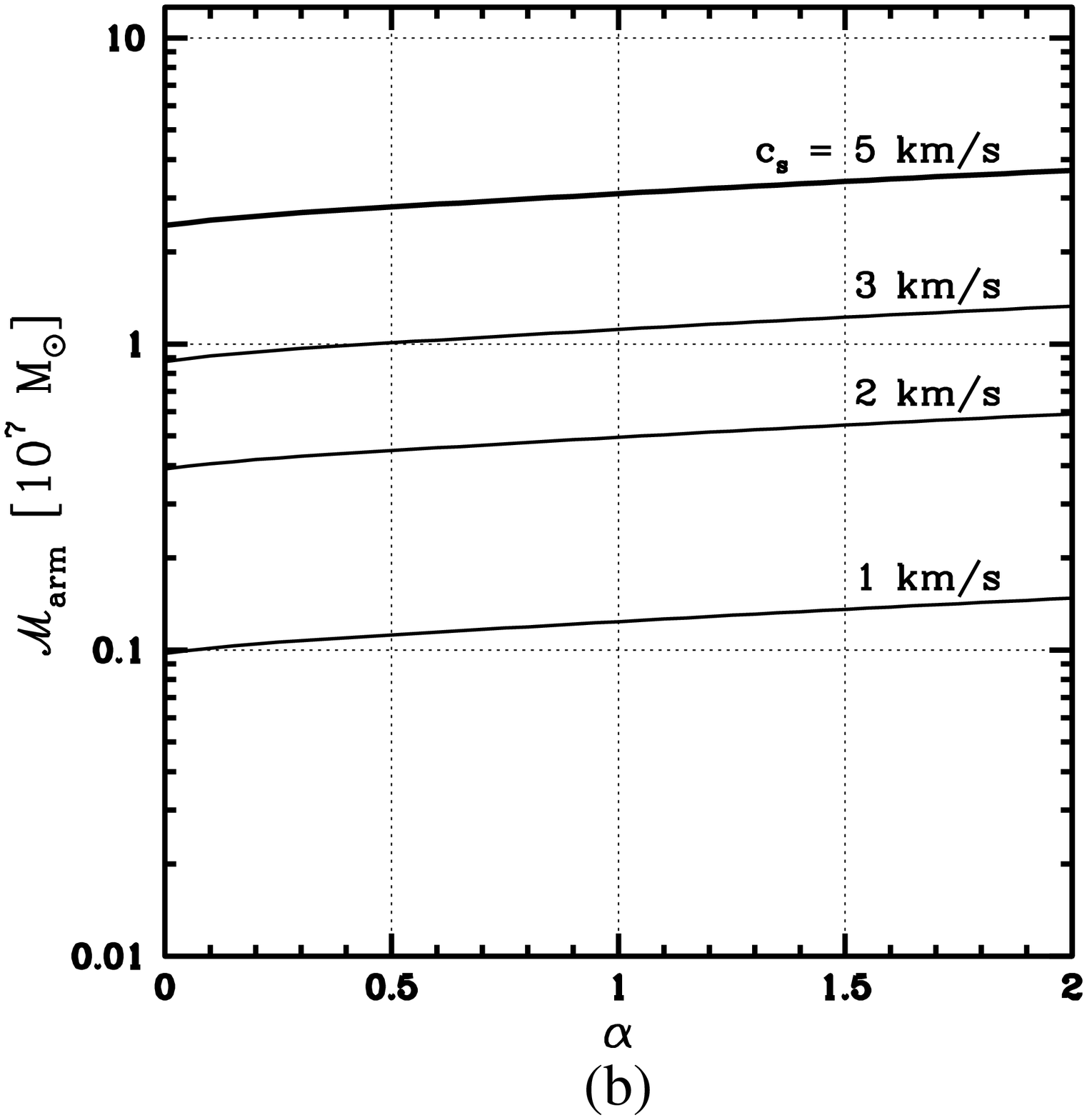}}
    \caption{Mass scales based on the critical wave number ({\it a}) and the
     spiral arm width ({\it b}) are plotted against magnetic-to-gas pressure
     ratio for four selected cases of velocity dispersion. Notice that the
     $n_{\rm o}\,\mu_{\rm ISM}$ factor is absorbed in the ordinate of ({\it a}).
     Once the global dynamics of rotating, self-gravitating, magnetized, gaseous
     disk fixes the arm width for us, the mass scale based on the arm width doesn't
     depend on the midplane density explicitly; it does on velocity dispersion
     to the second power. To prepare this figure $\Delta R$ = 1$\,$kpc is used.}
\end{figure}

\begin{figure}
  \figurenum{A1}
  \resizebox{\textwidth}{!}{\includegraphics{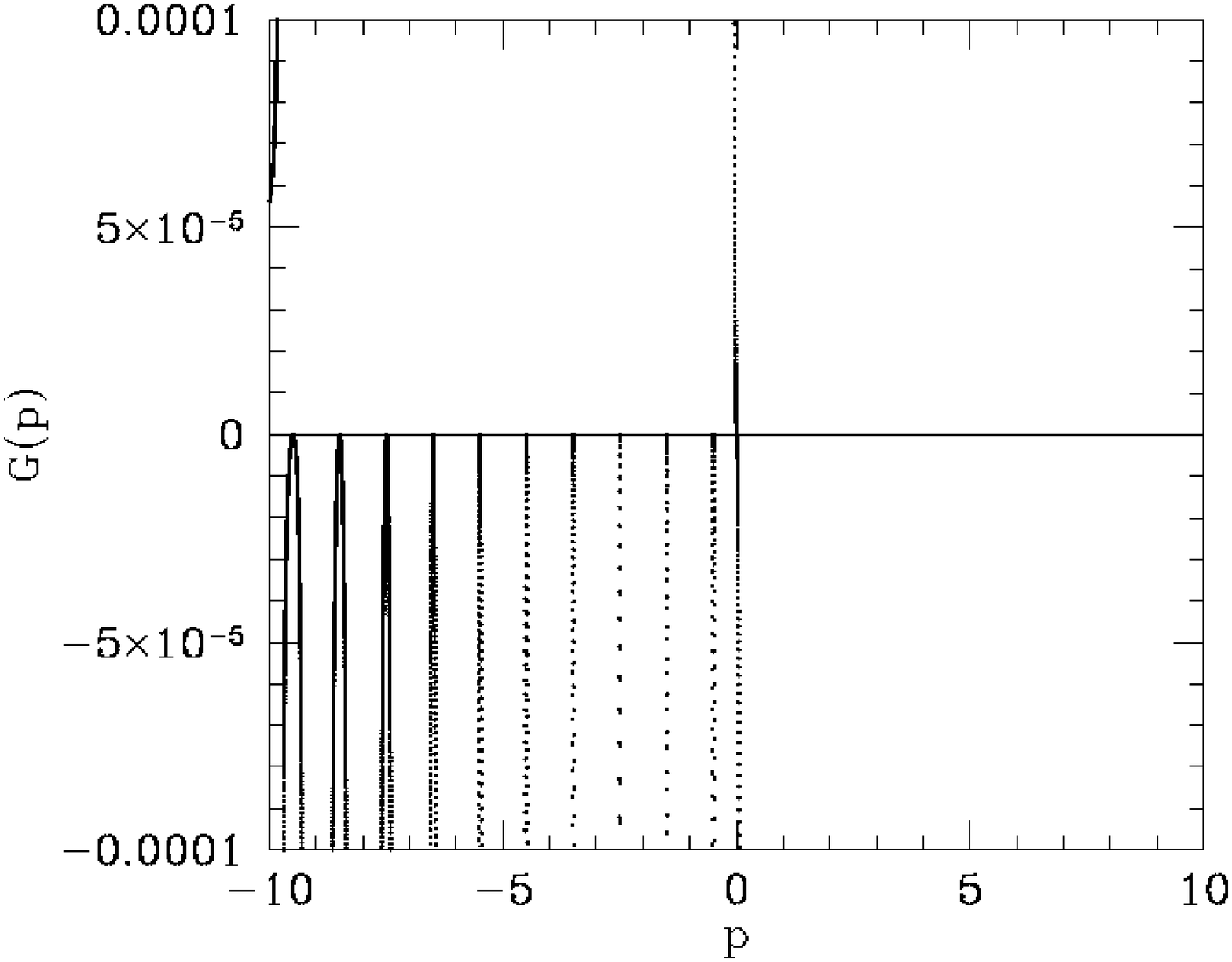}}
    \caption{$G(p\,)$ {\it versus} $p$ plot. The secular equation, B10,
    has zero solutions, where $p\,$ takes negative half integers including zero.}
\end{figure}

\begin{figure}
   \figurenum{A2}
  \resizebox{\textwidth}{!}{\includegraphics{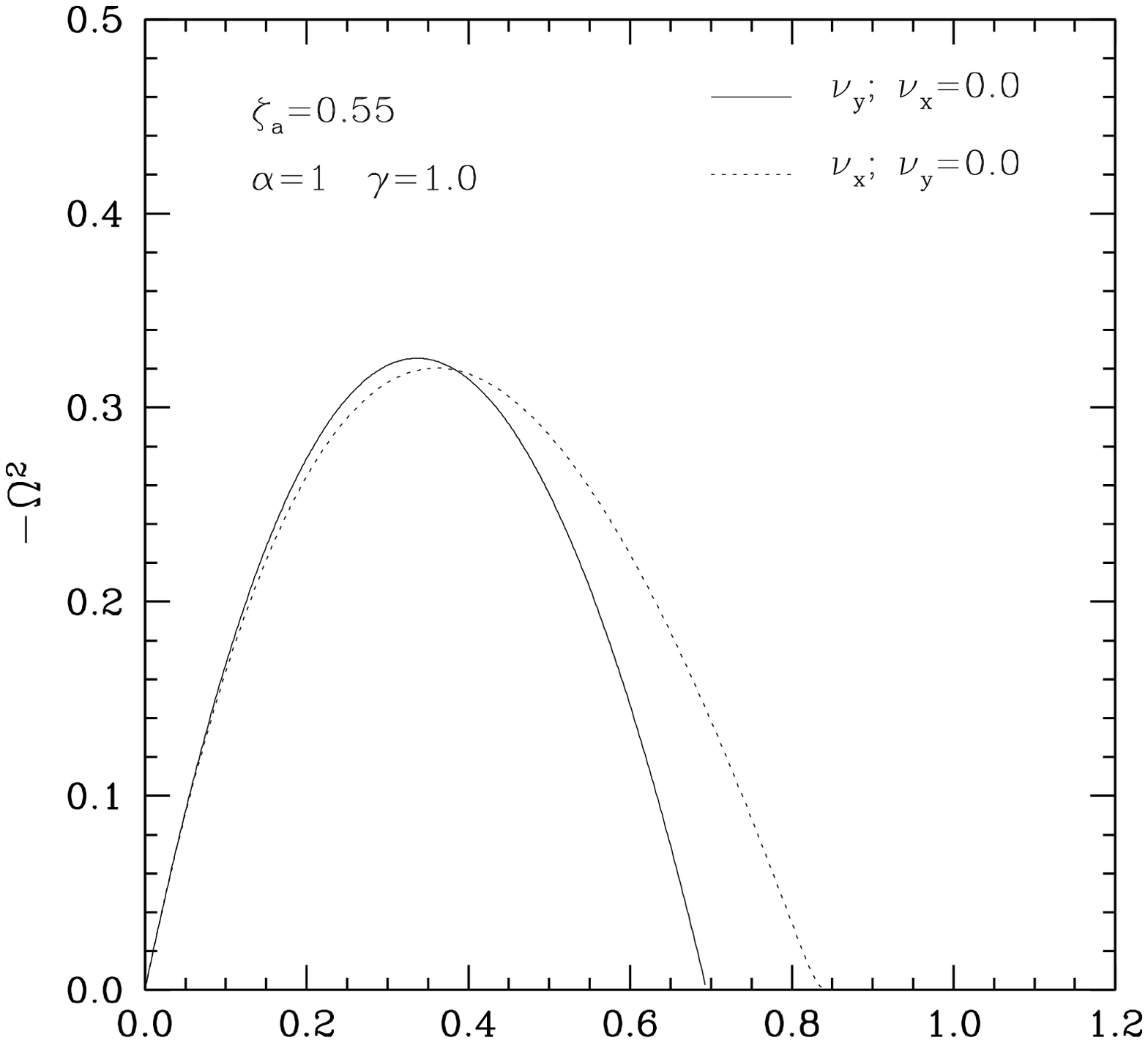}}

   \caption{Dispersion relations of the critically thick disk. The abscissa represents the normalized
    wave number $\nu_{y}$ for the undular mode (solid line), and $\nu_{x}$ for the interchange
    mode (dotted); while the ordinate denotes square of the normalized growth rate. Some of the
    parameters used in the calculation are given in the upper left corner of the frame.}
\end{figure}

\clearpage
\end{document}